\documentclass[%
 aip,
 jcp,   
 amsmath,amssymb,
 reprint,%
]{revtex4-1}

\usepackage{graphicx}
\usepackage{dcolumn}
\usepackage{bm}

\usepackage[utf8]{inputenc}
\usepackage[T1]{fontenc}
\usepackage{mathptmx}
\usepackage{etoolbox}

\makeatletter
\def\@email#1#2{%
 \endgroup
 \patchcmd{\titleblock@produce}
  {\frontmatter@RRAPformat}
  {\frontmatter@RRAPformat{\produce@RRAP{*#1\href{mailto:#2}{#2}}}\frontmatter@RRAPformat}
  {}{}
}%
\makeatother

\usepackage{color}

\DeclareMathAlphabet{\mathcal}{OMS}{cmsy}{m}{n}  

\newcommand*{\ntil}{\tilde n}
\newcommand*{\xtil}{\tilde x}
\newcommand*{\ncore}{n_{\rm core}}
\newcommand*{\ncoreA}{n_{\rm core}^{\cal A}}
\newcommand*{\xcoreA}{x_{\rm core}^{\cal A}}
\newcommand*{\ncoreB}{n_{\rm core}^{\cal B}}
\newcommand*{\xcoreB}{x_{\rm core}^{\cal B}}
\newcommand*{\ncoreC}{n_{\rm core}^{\cal C}}
\newcommand*{\xcoreC}{x_{\rm core}^{\cal C}}
\newcommand*{\nshell}{n_{\rm shell}}
\newcommand*{\nshellA}{n_{\rm shell}^{\cal A}}
\newcommand*{\xshellA}{x_{\rm shell}^{\cal A}}
\newcommand*{\nshellB}{n_{\rm shell}^{\cal B}}
\newcommand*{\xshellB}{x_{\rm shell}^{\cal B}}
\newcommand*{\nshellC}{n_{\rm shell}^{\cal C}}
\newcommand*{\xshellC}{x_{\rm shell}^{\cal C}}
\newcommand\ccb{{\cal A}}

\newcommand\ca{{\cal A}}
\newcommand\cb{{\cal B}}
\newcommand\cc{{\cal C}}
\newcommand\cg{{\cal G}}

\begin{document}


\title{Free energy surface of two-step nucleation}

\author{Dean Eaton}
\affiliation{Department of Physics, St. Francis Xavier University, Antigonish, NS, B2G 2W5, Canada}

\author{Ivan Saika-Voivod}
\affiliation{Department of Physics and Physical Oceanography, Memorial University of Newfoundland, \\St. John's, Newfoundland A1B 3X7, Canada}

\author{Richard K. Bowles}
\affiliation{Department of Chemistry, University of Saskatchewan, Saskatoon, SK, 57N 5C9, Canada}

\author{Peter H. Poole}
\email{ppoole@stfx.ca}
\affiliation{Department of Physics, St. Francis Xavier University, Antigonish, NS, B2G 2W5, Canada}

\begin{abstract}
We test the theoretical free energy surface (FES) for two-step nucleation (TSN) proposed by Iwamatsu [J. Chem. Phys. {\bf 134}, 164508 (2011)] by comparing the predictions of the theory to numerical results for the FES recently reported from Monte Carlo simulations of TSN in a simple lattice system [James, et al., J. Chem. Phys. {\bf 150}, 074501 (2019)].  No adjustable parameters are used to make this comparison.  That is, all the parameters of the theory are evaluated directly for the model system, yielding a predicted FES which we then compare to the FES obtained from simulations.
We find that the theoretical FES successfully predicts the numerically-evaluated FES over a range of thermodynamic conditions that spans distinct regimes of behavior associated with TSN.  All the qualitative features of the FES are captured by the theory and the quantitative comparison is also very good.
Our results demonstrate that Iwamatsu's extension of classical nucleation theory provides an excellent framework for understanding the thermodynamics of TSN.
\end{abstract}
\date{\today}
\maketitle

\section{Introduction}

In a simple nucleation process, the nucleus of a new stable phase forms and grows from within a homogeneous metastable phase, 
e.g.~when a liquid droplet appears in a supersaturated vapour. 
Classical nucleation theory (CNT) has long been a valuable tool for conceptualizing the nature of simple nucleation processes and for providing a starting point for quantitative estimates of nucleation barriers and rates~\cite{Debenedetti:1996,Kashchiev:2000,Kelton:2010}.  The accuracy and limitations of CNT have been tested by comparisons with experiments and computer simulations.  In the case of simulations, tests of CNT are often facilitated by the direct evaluation of the physical parameters that appear in the theory, such as chemical potentials, surface tensions, and nucleus growth rates.
Such tests have guided the refinement of CNT-based approaches to improve quantitative predictions for real systems;  see e.g. Refs.~\onlinecite{Ryu:2010gw,Espinosa:2014di,Richard:2018ee}.

\begin{figure}[t]
\centerline{\includegraphics[scale=0.35]{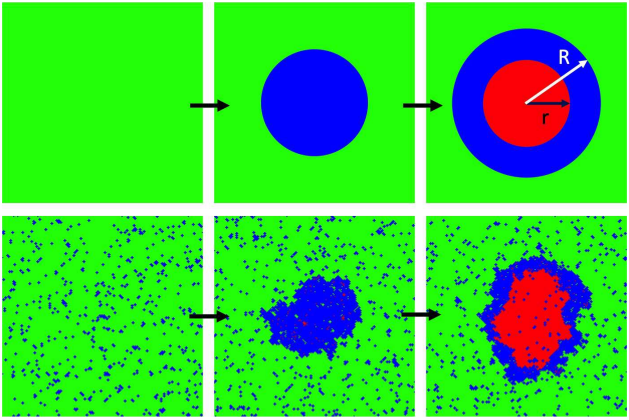}}
\caption{Upper panels:  Schematic depiction of TSN.  The bulk metastable $\ca$ phase (green) is initially homogeneous (left panel).  A fluctuation of the intermediate $\cb$ phase (blue) then appears (middle panel).  Finally, the stable $\cc$ phase (red) appears within the finite-sized region of the $\cb$ phase (right panel).  Lower panels:  Configurations from $L=200$ umbrella sampling simulations of the 2D metamagnet at $H=3.96$ and $H_s=0.01$ showing the corresponding stages by which 
the bulk metastable $\ca$ phase (left panel) develops a fluctuation of the $\cb$ phase (middle panel), within which the nucleus of the $\cc$ phase then forms (right panel).}
\label{fig1}
\end{figure}

Complex non-classical nucleation processes that deviate significantly from the predictions of CNT are receiving increased attention in recent years~\cite{Sosso:2016bda,Karthika:2016jj,Jehannin:2019cq,
Zhou:2019gh}.  
In particular, ``two-step nucleation" (TSN) has become a focus of interest
due to its role in important phenomena such as biomineralization and protein crystallization~\cite{Vekilov:2004jc,vanMeel:2008hb,Vekilov:2010gm,Iwamatsu:2011if,Sear:2012ji,Qi:2015iea,Lutsko:2019hja,Kashchiev:2020iz,Lvov:2020kr,Shao:2020ej}.
The TSN process is shown schematically in the upper panels of Fig.~\ref{fig1}.  
In the first step of TSN, a cluster of an intermediate phase appears within the homogeneous metastable phase.
In the second step, the stable phase appears and grows from within the finite cluster of the intermediate phase.  
TSN has been identified and studied in a growing range of systems, both in experiments~\cite{Vekilov:2004jc,Vekilov:2010gm,Peng:2014is,Qi:2015iea,Ishizuka:2016ep,Zhang:2017ju,Yamazaki:2017jb,Li:2008js,Pouget:2009dg,Ou:2019ff,Fang:2020cq}
and simulations~\cite{Duff:2009p6360,Vatamanu:2010ixa,Whitelam:2010dr,Wallace:2013df,Qi:2015iea,Lifanov:2016dua,PhysRevE.98.032606,Kumar:2018go,James:2019gi,Schmid:2019gka,Shi:2019daa,Jiang:2019ez,Lee:2019bq,Arjun:2019fl}.

To develop a theoretical framework for TSN it is natural to use CNT as a starting point~\cite{Iwamatsu:2011if,Qi:2015iea,Banerjee:2018ci,Kashchiev:2020iz,Shao:2020ej}.
The thermodynamics of simple (i.e. one-step) nucleation in CNT is described by the model for the free energy of formation of a nucleus of size $n$, which in three dimensions is given by,
\begin{equation}
G=
n\, \Delta \mu+\phi\, n^{2/3} \sigma.
\label{cnt1d}
\end{equation}
Here $\Delta \mu$ is the difference in chemical potential between the metastable and stable phases, $\sigma$ is the surface tension between the two phases, and $\phi$ is a shape factor.  This model for $G$ captures the competition between the decrease of the free energy as $n$ monomers coalesce to form a cluster of the stable phase, and the increase in the free energy due to the cost of the interface, having surface area $\phi\, n^{2/3}$, separating the two phases.  In the case of TSN, the reaction coordinate $n$ must be replaced by at least two coordinates, e.g. one to quantify the overall size of the nucleus, and another to specify the proportion of the intermediate and stable phases occurring within the nucleus.  The model for $G$ in TSN will therefore be a free energy surface (FES), rather than a single-variable function such as that given in Eq.~\ref{cnt1d}. 

An early proposal for the FES of TSN was presented by Iwamatsu in 2011~\cite{Iwamatsu:2011if}.  As described in detail below, this model FES is formulated as a sum of two CNT-like contributions, one for the formation of the intermediate phase from the metastable phase, and the other for the formation of the stable phase within the intermediate phase droplet.  An additional term is included to account for the interaction between the interfaces within the multiphase nucleus.  Variations of this model have been studied in subsequent work, which have shown it to be successful in predicting qualitative phenonmena characteristic of TSN~\cite{Iwamatsu:2011if,Qi:2015iea,Kashchiev:2020iz,Shao:2020ej}.  
We note that expressions similar to that in Ref.~\onlinecite{Iwamatsu:2011if} had previously been used to describe the related phenomena of deliquescence and efflorescence, which require consideration of the free energy of a multiphase droplet surrounded by a metastable vapor phase~\cite{Djikaev:2011p7210,Shchekin:2013ev}.

Despite the interest in using a CNT-inspired approach to model TSN, there have been comparatively few studies which quantitatively test the predictions of a proposed FES against results obtained from experiments or simulations.  Ref.~\onlinecite{Qi:2015iea} shows that an expression for the FES similar to that proposed in Ref.~\onlinecite{Iwamatsu:2011if} 
compares well to the FES found from simulations of TSN in a quasi-2D colloidal system.  The success of this comparison suggests that a more extensive test is warranted.  
In a recent study of TSN in a 2D Ising-like lattice model, Ref.~\onlinecite{James:2019gi} presents high-resolution results for the FES obtained from simulations over a wide range of thermodynamic conditions.  These results are well-suited for comparison with an analytic theory.  Accordingly, the goal of the present work is to use the simulation results of Ref.~\onlinecite{James:2019gi} to test the theory for the FES of TSN proposed in Ref.~\onlinecite{Iwamatsu:2011if}.

As shown below, we conduct this test by first evaluating all the required parameters of the theory for the FES from the model system itself.
Ref.~\onlinecite{James:2019gi} 
already provides the required data for the chemical potentials and surface tensions of the bulk phases involved in the observed TSN process.  In the present work, we separately calculate the parameters required to model the interaction between the two interfaces occurring in the multiphase nucleus.
As a result, we are able to present a comparison of the FES as predicted by theory and as obtained directly from simulations that does not depend on any adjustable parameters.  

This paper is organized as follows:  Section~II describes Iwamatsu's model for the FES of TSN~\cite{Iwamatsu:2011if} 
and re-expresses it in a form appropriate for comparison with the results of Ref.~\onlinecite{James:2019gi}.  Section~III describes the lattice model studied in Ref.~\onlinecite{James:2019gi} and Section~IV summarizes the simulation results for the FES and other thermodynamic properties calculated in Ref.~\onlinecite{James:2019gi}.  Section~V presents new simulations to determine the interaction parameters required to model the interaction of two nearby interfaces.  A comparison of the predicted and simulated FES is given in Section~VI, followed by a discussion in Section VII.  The Supplemental Materials (SM) provide details on the order parameters used to characterize the FES and additional information on the simulations methods used to calculate the system free energy.

\section{Theory for the free energy surface}

The upper panels of Fig.~\ref{fig1} show an idealized TSN process in which the end product is 
a nucleus having a core-shell structure.  Surrounded by the bulk metastable phase $\ca$ (green), this nucleus has an outer shell of the intermediate $\cb$ phase (blue) and a core region of the stable $\cc$ phase (red).
Ref.~\onlinecite{Iwamatsu:2011if} expresses the free energy to create a nucleus having a core-shell structure as the sum of three contributions,
\begin{equation}
G=G_{\ca\cb}(R)+G_{\cb\cc}(r)+G_{\rm int}.
\end{equation}
Here, $G_{\ca\cb}(R)$ 
is the free energy to create a region of phase $\cb$ of radius $R$ within phase $\ca$.
$G_{\cb\cc}(r)$ 
is the free energy to create a core region of phase $\cc$ of radius $r$ within the $\cb$ region. $G_{\rm int}$ models the contribution to the free energy resulting from the interaction of the $\ca\cb$ and $\cb\cc$ interfaces.

Ref.~\onlinecite{Iwamatsu:2011if} uses CNT expressions for $G_{\ca\cb}(R)$ and $G_{\cb\cc}(r)$ of the form of Eq.~\ref{cnt1d} to express $G$ for a three dimensional core-shell nucleus as,  
\begin{eqnarray}
G&=&
n\, \Delta \mu_{\ca\cb} +4\pi R^2\, \sigma_{\ca\cb} \cr
&+&n_{\rm core}\, \Delta \mu_{\cb\cc} +4\pi r^2\, \sigma_{\cb\cc} \cr
&+&4\pi R^2 S\exp\big [-(R-r)/\xi\big ].
\label{iwa}
\end{eqnarray}
In Eq.~\ref{iwa}, the total number of particles in the nucleus (core and shell) is $n$, and the core contains $n_{\rm core}$ particles.  $\Delta \mu_{\ca\cb}=\mu_\cb-\mu_\ca$ is the difference in the chemical potential between the bulk phases $\cb$ and $\ca$, and $\sigma_{\ca\cb}$ is the $\ca\cb$ surface tension.  $\Delta \mu_{\cb\cc}$ and $\sigma_{\cb\cc}$ are similarly defined.  
The last term in Eq.~\ref{iwa} models $G_{\rm int}$ and is related to the disjoining pressure associated with the double interface~\cite{Djikaev:2011p7210,Iwamatsu:2011if}.
The spreading parameter $S$ is defined by,
\begin{equation}
S=\sigma_{\ca\cc} - \sigma_{\ca\cb} - \sigma_{\cb\cc},
\end{equation}
where $\sigma_{\ca\cc}$ 
is the $\ca\cc$ surface tension.
Note that $\sigma_{\ca\cc}$ is the surface tension of an interface where the $\ca$ and $\cc$ phases are in direct contact, without a wetting layer of $\cb$ between them.
The length scale $\xi$ characterizes the range of the interaction between the $\ca\cb$ and $\cb\cc$ interfaces.
A nucleus morphology that conforms to the core-shell structure depicted in Fig.~\ref{fig1} requires that $S\ge 0$, which corresponds to a repulsive interaction between the $\ca\cb$ and $\cb\cc$ interfaces, and complete wetting of the $\cc$ phase by the $\cb$ phase.
Eq.~\ref{iwa} reduces to the conventional CNT expression for direct (i.e. one-step) nucleation from $\ca$ to $\cc$ when $r=R$, in which case the $\cb$ phase never appears 
as an intermediate phase or as a
wetting layer during the nucleation process.

To generalize our analysis to both two and three dimensions, we write 
the surface area $A$ of a cluster of $n$ particles as $A=\phi\, n^\alpha$, where $\alpha=(D-1)/D$ depends on the dimension of space $D$.  We further write the radius $R$ of a cluster of size $n$ as $R=\psi\,n^\gamma$, where $\gamma=1/D$.  
For a circular cluster in $D=2$, we have $\alpha=\gamma=1/2$, $\phi=(4\pi v)^{1/2}$ and $\psi=(v/\pi)^{1/2}$,
where $v$ is the area per molecule.
For a spherical cluster in $D=3$, we have $\alpha=2/3$, $\gamma=1/3$, $\phi=(36\pi v^2)^{1/3}$ 
and $\psi=(3v/4\pi)^{1/3}$,
where $v$ is the volume per molecule.  

We define the composition of the nucleus as,
\begin{equation}
 x=\ncore/n.
\label{defx}
\end{equation}
With the above definitions, we can rewrite the FES described by Eq.~\ref{iwa} solely in terms of $n$ and $x$ as,
\begin{eqnarray}
G(n,x) 
&=& n\, \Delta \mu_{\ca\cb} + \phi\, n^\alpha\, \sigma_{\ca\cb} \cr
&+&xn\, \Delta \mu_{\cb\cc} +\phi\, (xn)^\alpha\, \sigma_{\cb\cc} \cr
&+&\phi\, n^\alpha S \exp\big [-\psi\, n^\gamma(1-x^\gamma)/\xi\big ].
\label{ours}
\end{eqnarray}
Eq.~\ref{ours} expresses the model of Ref.~\onlinecite{Iwamatsu:2011if} for the FES for TSN in a form that can be directly compared with simulation results for the FES obtained in terms of $n$ and $x$. These order parameters have been chosen in previous studies of the FES for TSN~\cite{Duff:2009p6360,Schmid:2019gka} including Ref.~\onlinecite{James:2019gi}.


\section{Lattice model and cluster properties}

We test Eq.~\ref{ours} using the results obtained for the lattice model described in detail in Ref.~\onlinecite{James:2019gi}.  This system is a $D=2$ model of a metamagnet in which Ising spins $s_i=\pm 1$ interact via antiferromagnetic nearest-neighbor (nn) and ferromagnetic next-nearest-neighbour (nnn) interactions on a square lattice of $N=L^2$ sites with periodic boundary conditions~\cite{Landau:1972mt,Landau:1981bi,Rikvold:1983fs,Herrmann:1984fs}.
The energy $E$ of a microstate is,
\begin{equation}
\frac{E}{J} = \sum_{\langle \rm nn \rangle} s_i s_j
- \frac{1}{2} \sum_{\langle \rm nnn \rangle} s_i s_j 
 -H\sum_{i=1}^N s_i
-H_s\sum_{i=1}^N \sigma_i s_i,
\label{ham}
\end{equation}
where $J$ is the magnitude of the nn interaction energy.  
$H$ is the direct magnetic field, $H_s$ is the staggered field, and $\sigma_i=(-1)^{x_i+y_i}$, 
where $x_i$ and $y_i$ are respectively the integer horizontal and vertical coordinates of site $i$.

\begin{figure}
\centerline{\includegraphics[scale=0.35]{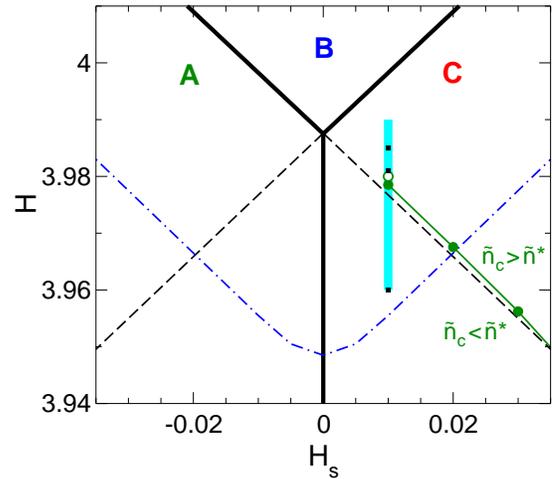}}
\caption{Phase diagram of the metamagnet for $kT/J=1$ in the vicinity of the $\ca\cb\cc$ triple point. Solid black lines are coexistence lines and dashed black lines are metastable extensions of coexistence lines.  The vertical cyan bar shows the range of states at $H_s=0.01$ that we focus on this work. Small black squares locate the three states studied in Fig.~\ref{g3d}.
The blue dotted-dashed line is the limit of metastability of the bulk $\cb$ phase for a system of size $L = 64$. Filled green circles locate points on the line at which $\ntil_c=\ntil^*$ as obtained from MC simulations.  The green open circle is the point at which $\ntil_c=\ntil^*$ at $H_s=0.01$ as predicted by Eq.~\ref{ours}.}
\label{pd}
\end{figure}

Ref.~\onlinecite{James:2019gi} 
studied this metamagnet model at a fixed temperature $T=J/k$, where $k$ is Boltzmann's constant.  At fixed $T=J/k$ and for $H>0$, this system may be found in one of three phases: 
an antiferromagnetic phase
$\ca$ having a ground state at $T=0$ with all $s_i=-\sigma_i$;
a ferromagnetic phase
$\cb$ having a ground state at $T=0$ with all $s_i=1$; and
an antiferromagnetic phase
$\cc$ having a ground state at $T=0$ with all $s_i=\sigma_i$.
As shown in Fig.~\ref{pd}, the phase diagram of the system in the plane of $H_s$ and $H$ contains three coexistence lines, one for each pair of the three phases, which meet at a triple point at $(H^T_s,H^T)=(0,3.9876)$.  

\begin{figure*}
\newcommand\x{0.55}
{\includegraphics[scale=\x]{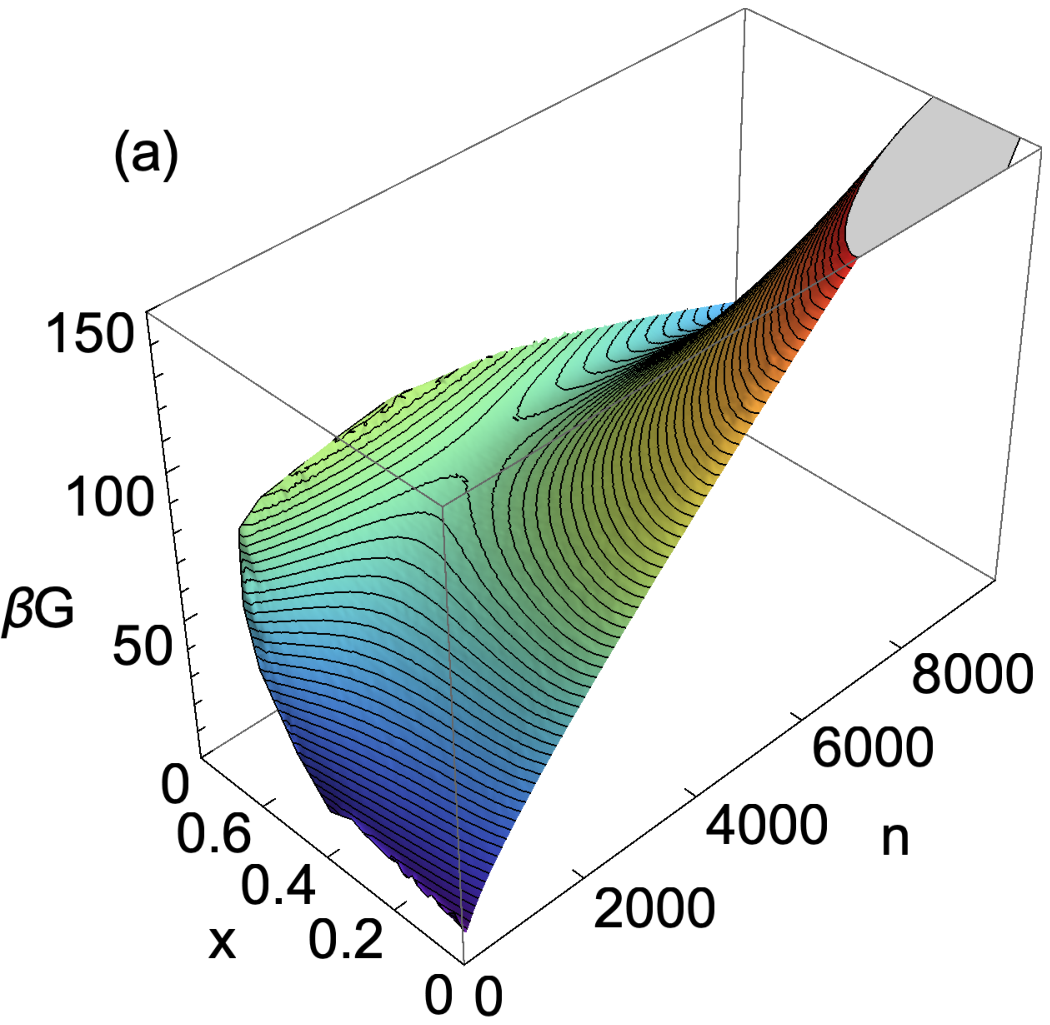}}
{\includegraphics[scale=\x]{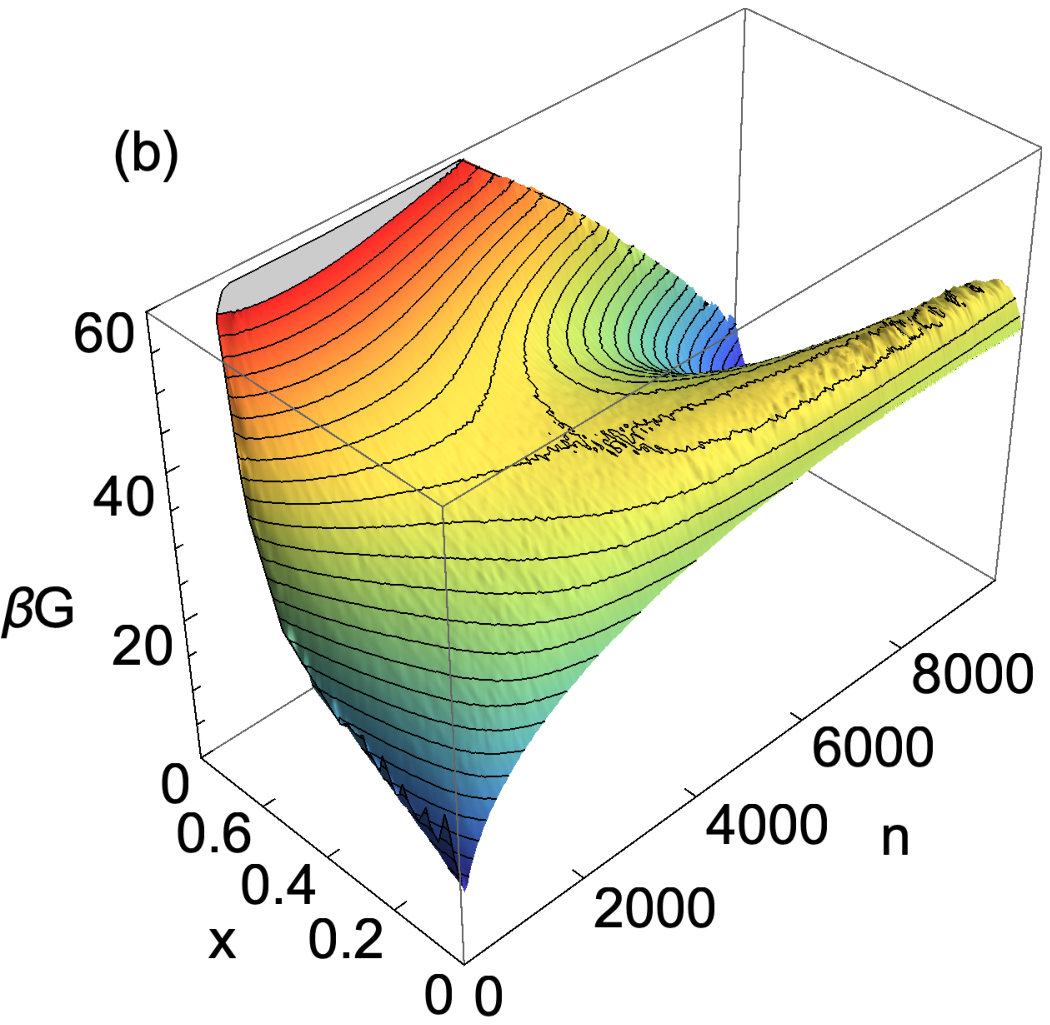}}
{\includegraphics[scale=\x]{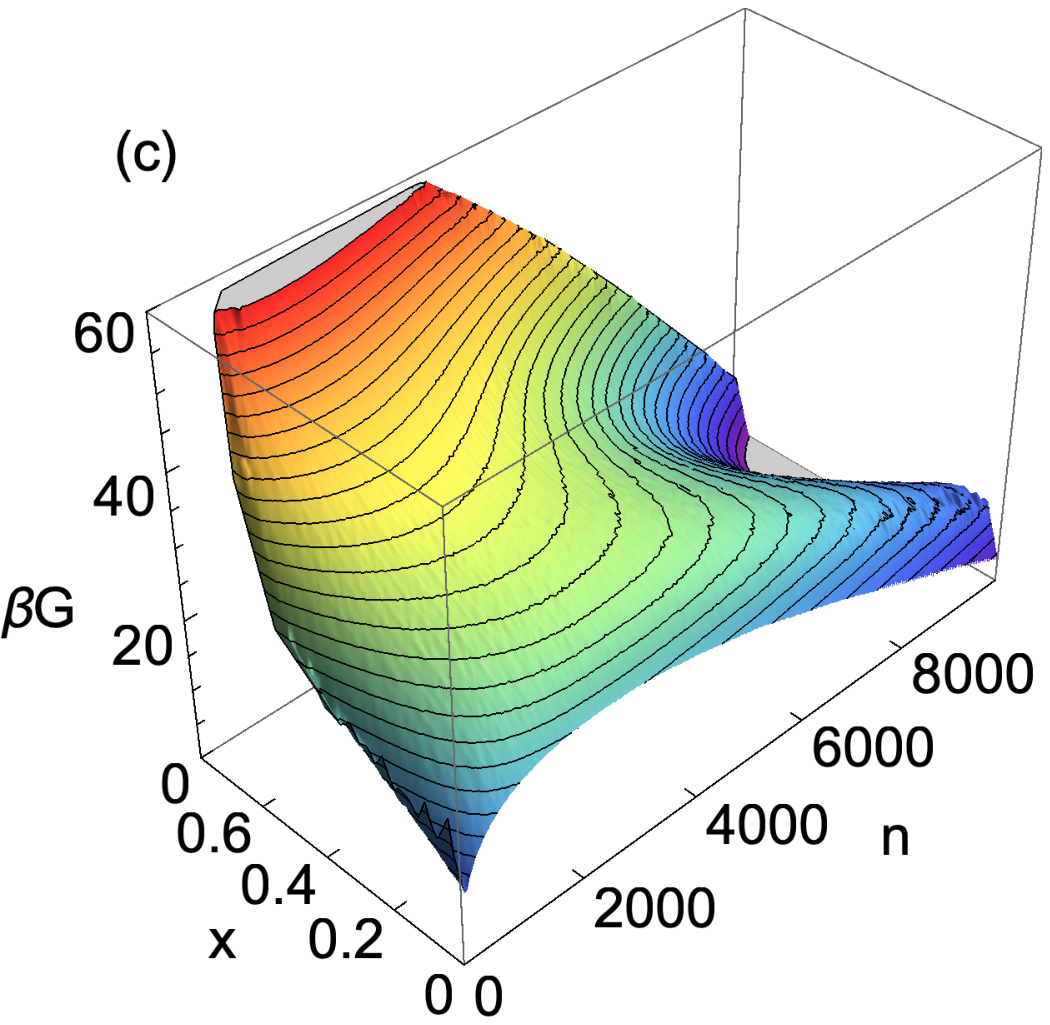}}
{\includegraphics[scale=\x]{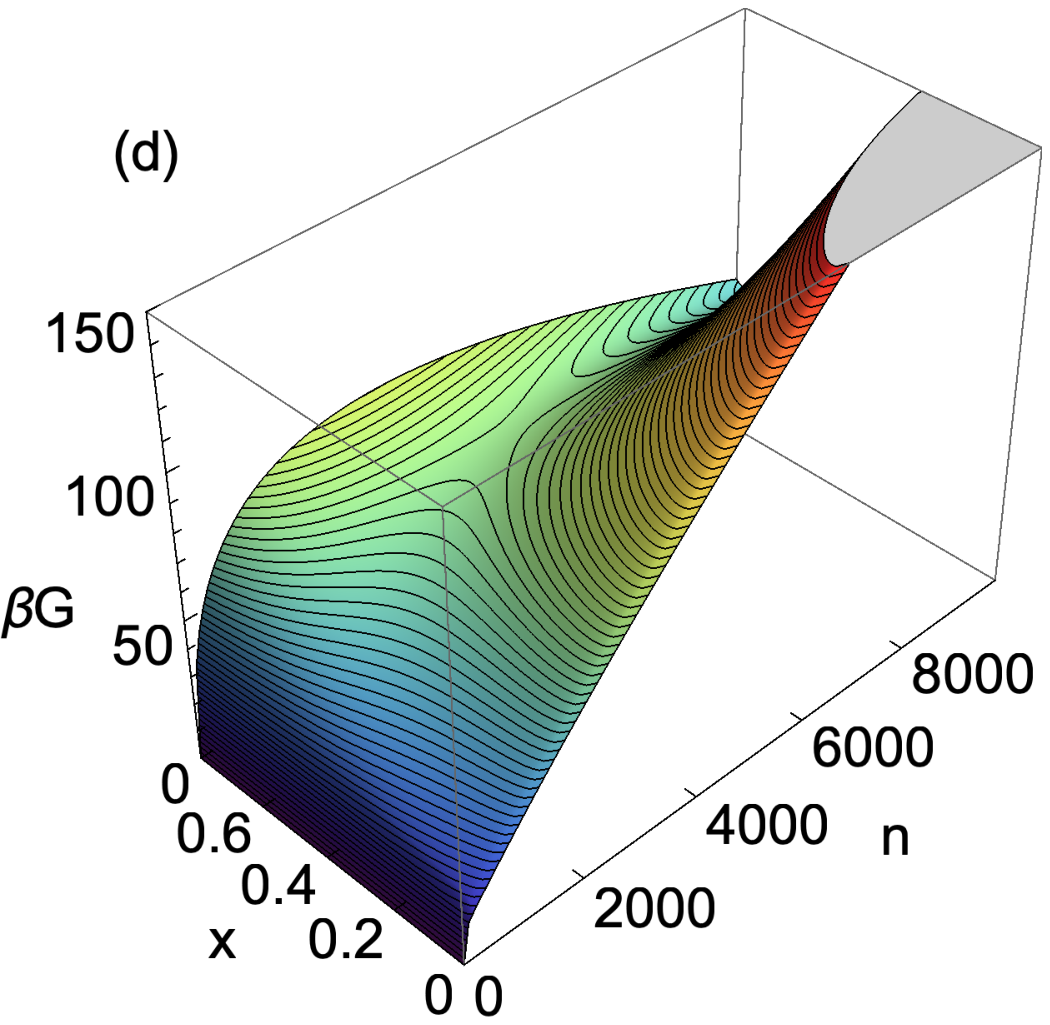}}
{\includegraphics[scale=\x]{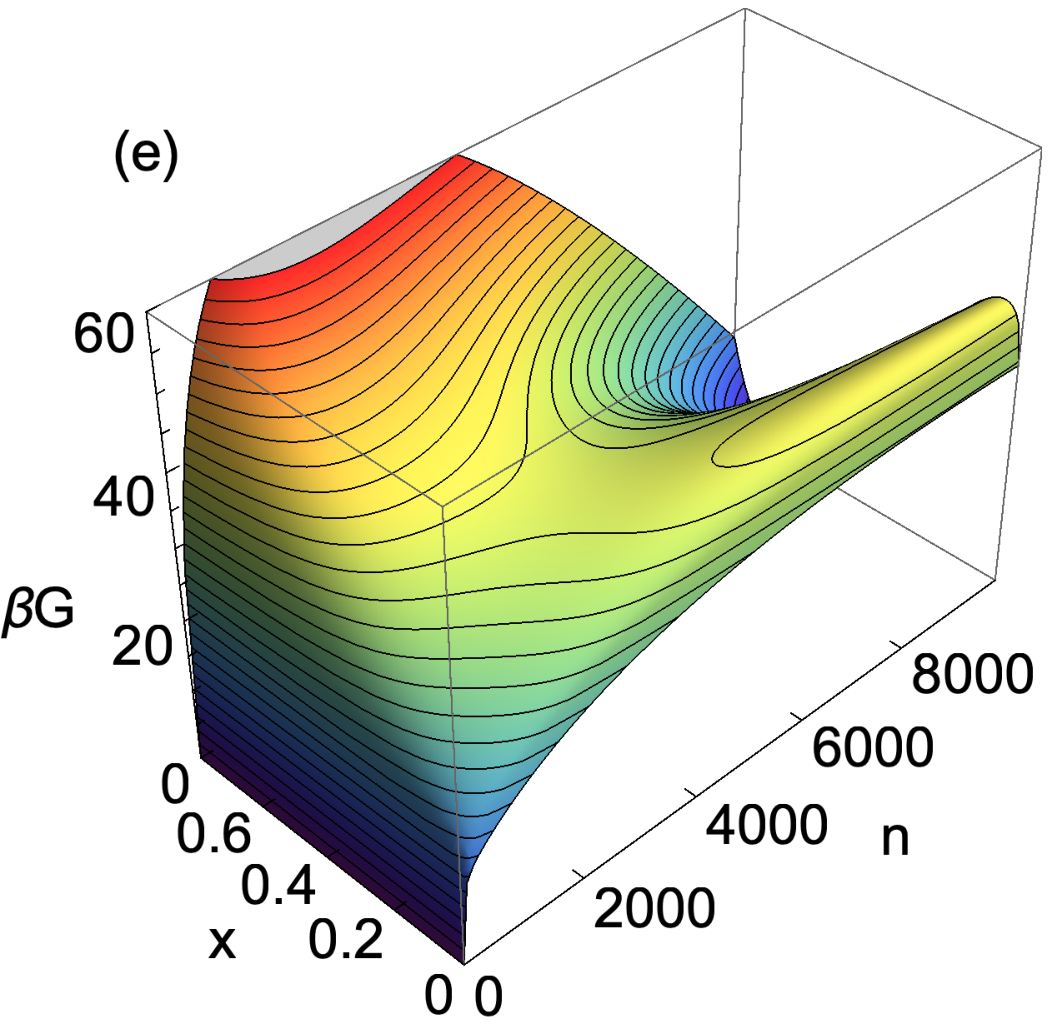}}
{\includegraphics[scale=\x]{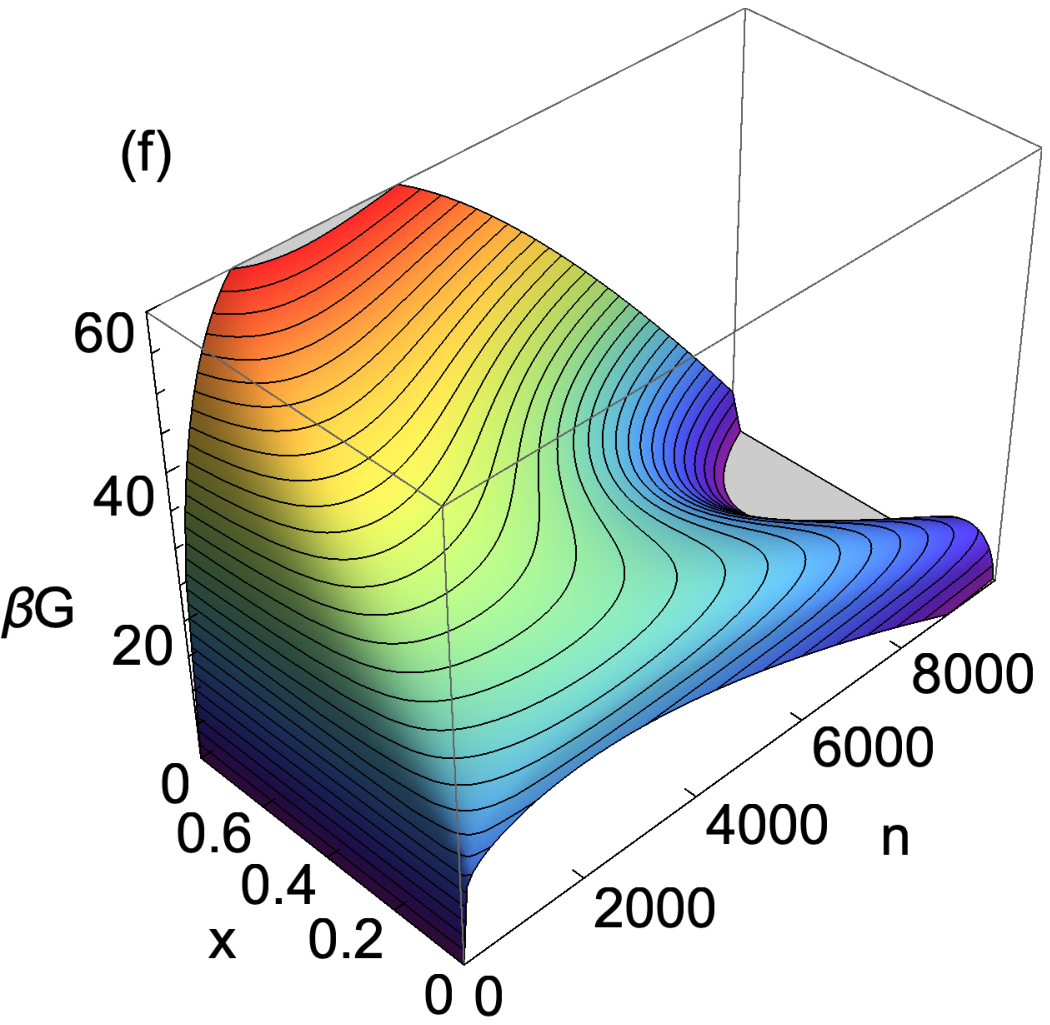}}
\caption{Comparison of $G(n,x)$ as found from MC simulations (upper panels) and as predicted by Eq.~\ref{ours} (lower panels).
Upper panels show $G(n,x)$ as computed from the MC simulation data presented in Ref.~\onlinecite{James:2019gi} 
for $H_s=0.01$ and for $H=\{3.96, 3.981, 3.985\}$ in (a,b,c) respectively.  The location of these three states are marked in the phase diagram in Fig.~\ref{pd} by small black squares.
The error in $G(n,x)$ is typically less than $1kT$.
Lower panels show $G(n,x)$ as predicted by Eq.~\ref{ours} for the same three states as in the upper panels. That is, $H_s=0.01$ and $H=\{3.96, 3.981, 3.985\}$ in (d,e,f) respectively.  
In all panels, contours are $2kT$ apart.}
\label{g3d}
\end{figure*}

\begin{figure}
{\includegraphics[scale=0.6]{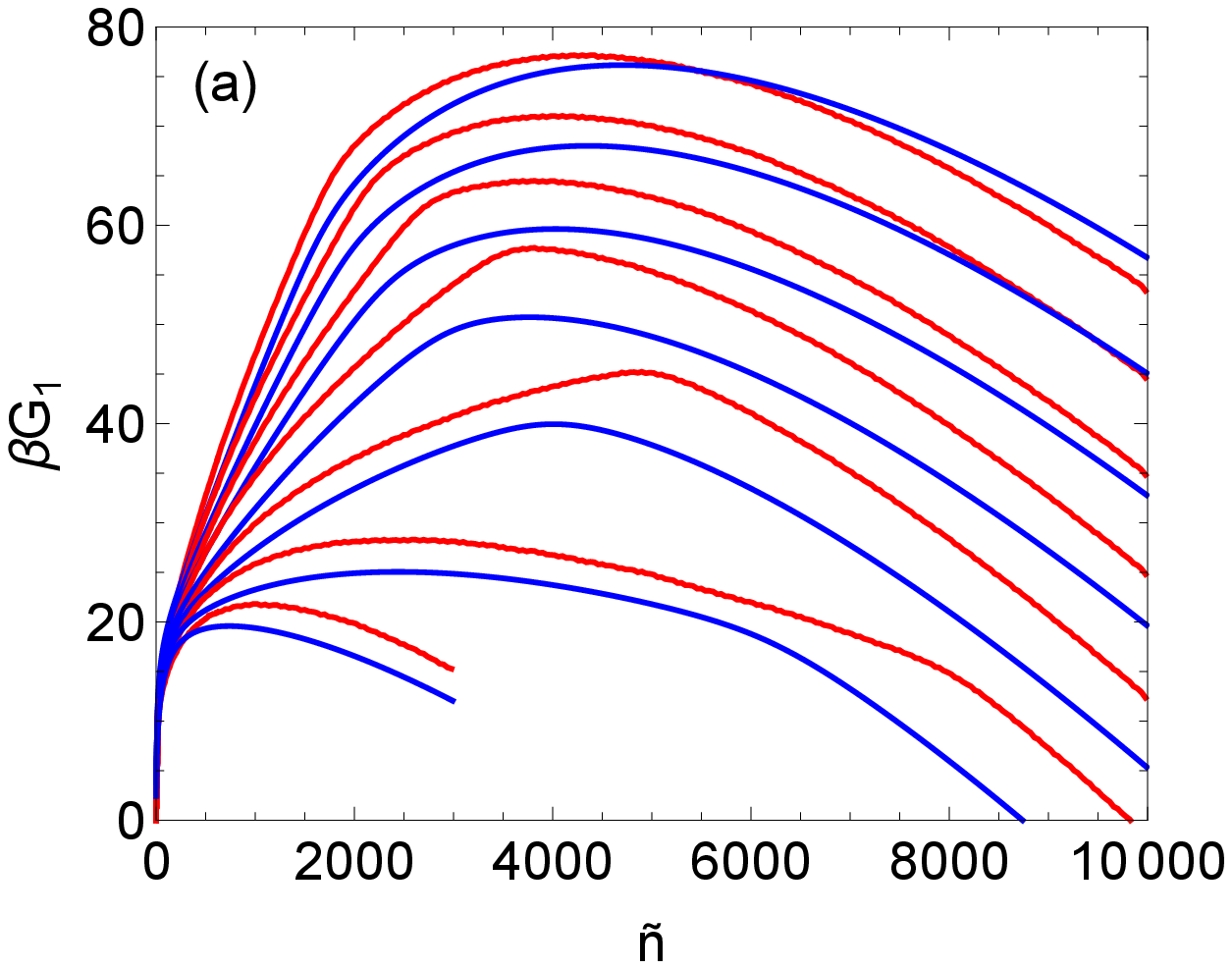}}
{\includegraphics[scale=0.6]{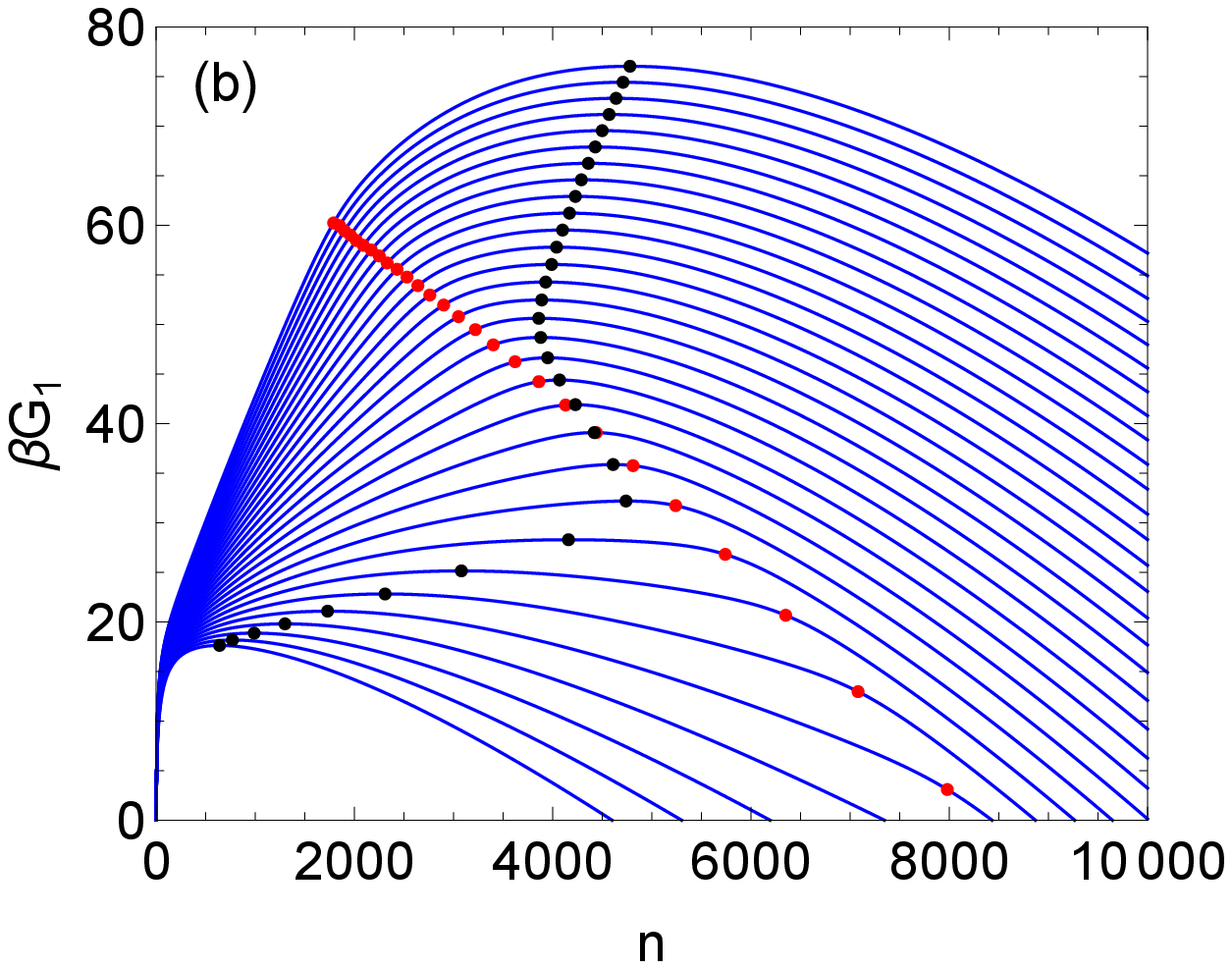}}
\caption{
(a) $G_1(\ntil)$ as measured in MC simulations (red) and as predicted by Eqs.~\ref{ours} and \ref{g1} (blue) for $H_s=0.01$ and $H=3.96$ to $3.99$ in steps of 0.005 from top to bottom.
(b) $G_1(n)$ from Eqs.~\ref{ours} and \ref{g1} for $H_s=0.01$ and $H=3.96$ to $3.99$ in steps of $0.001$ from top to bottom.  For each curve, the dots locate the values of $n_c$ (red) and $n^*$ (black).
}
\label{gnon}
\end{figure}

\begin{figure}
{\includegraphics[scale=0.6]{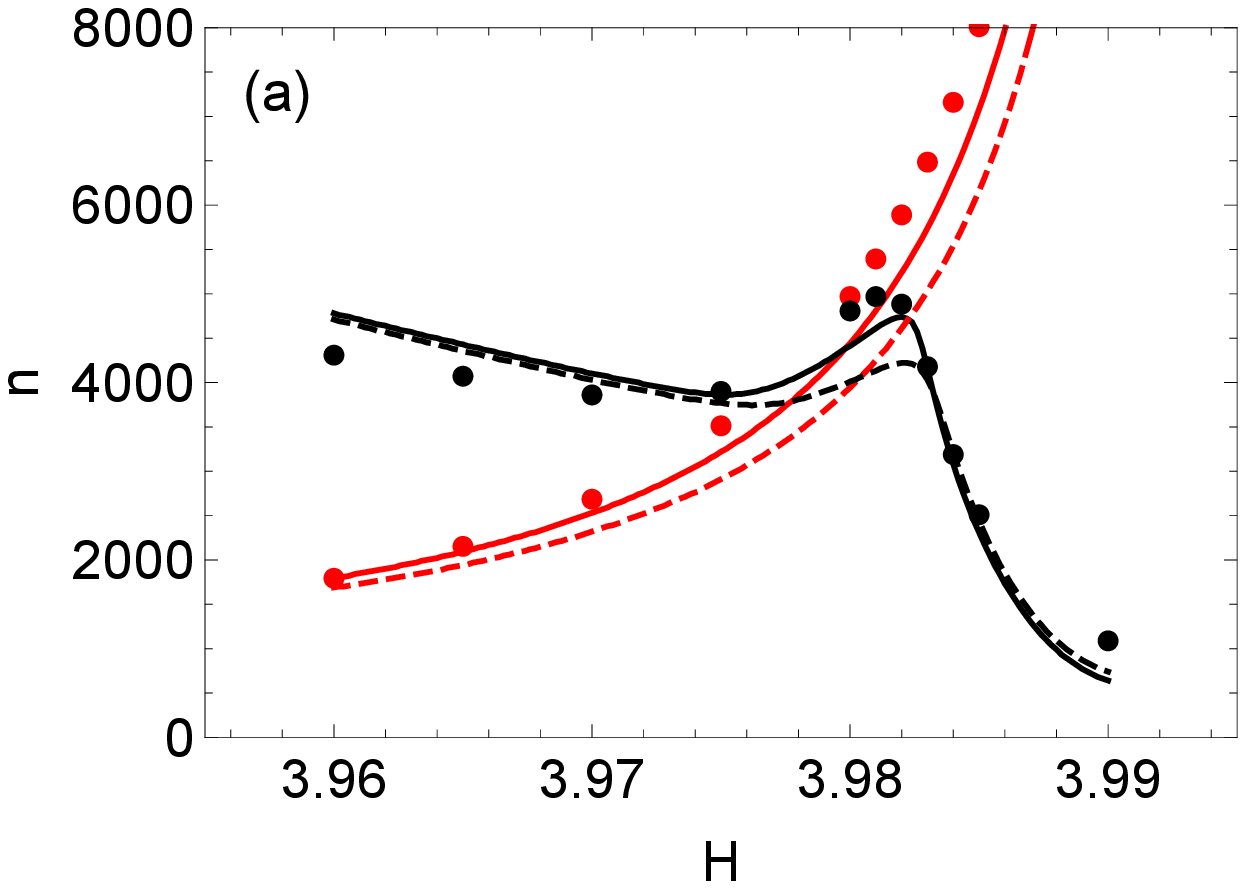}}
{\includegraphics[scale=0.6]{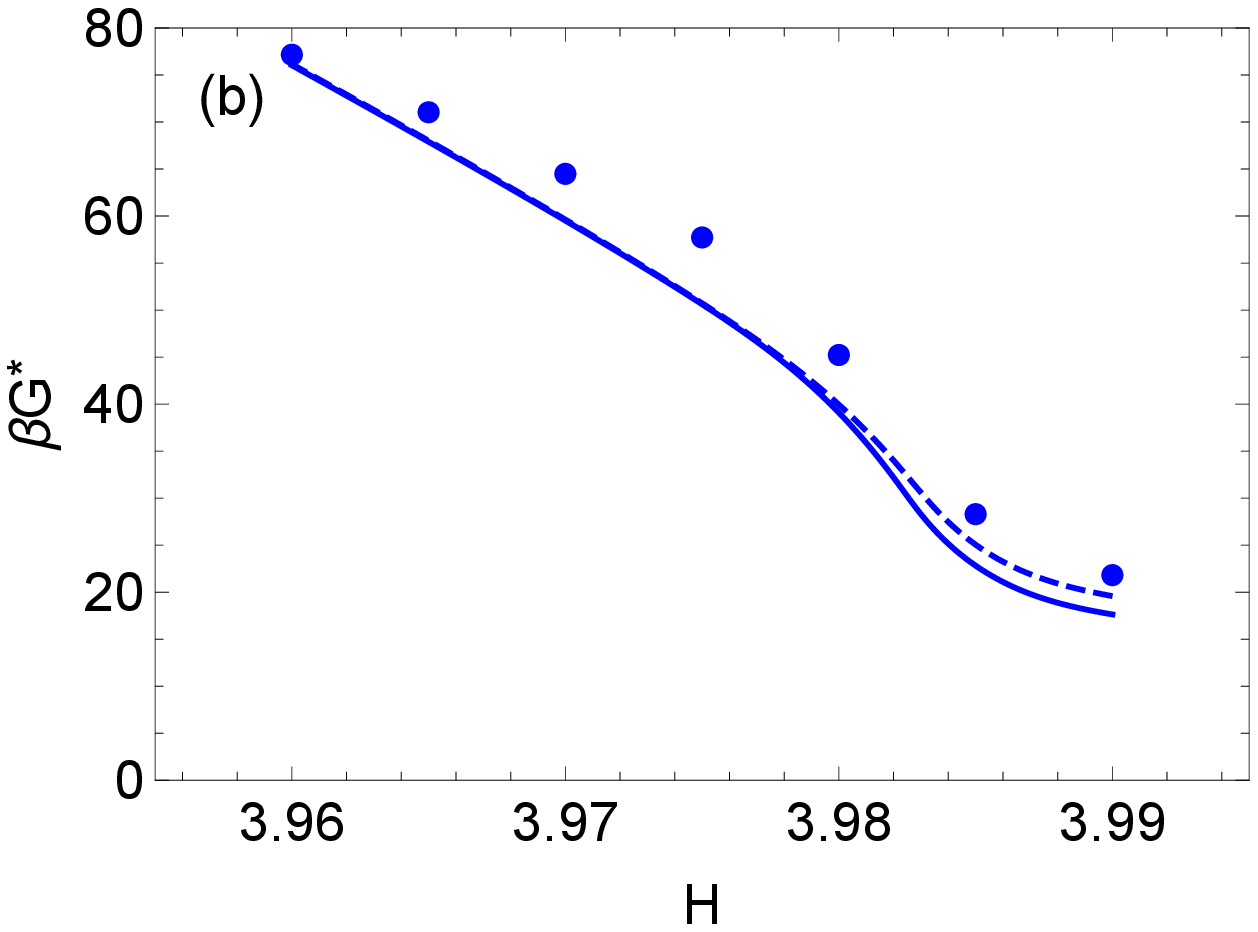}}
\caption{(a) Circles show $\ntil^*$ (black) and $\ntil_c$ (red) as found from MC simulations.  Solid lines are model predictions for 
$n^*$ (black) and $n_c$ (red).
Dashed lines are model predictions for 
$\ntil^*$ (black) and $\ntil_c$ (red).
(b) $G^*$ versus $H$.  Circles are found from $G_1(\ntil)$ as evaluated in MC simulations.  Also shown are the model predictions for $G^*$ as determined from $G_1(n)$ (solid line) and $G_1(\ntil)$ (dashed line).}
\label{gstar}
\end{figure}

Ref.~\onlinecite{James:2019gi} 
uses Monte Carlo simulations to study the TSN process 
in which the metastable $\ca$ phase transforms to the stable $\cc$ phase, 
during which the $\cb$ phase plays an intermediate role in the nucleation process.
As described in Ref.~\onlinecite{James:2019gi}, it is straightforward to identify local clusters of sites belonging to the 
$\ca$, $\cb$ and $\cc$ 
phases in any given configuration of the lattice model.  An example is shown in the lower panels of Fig.~\ref{fig1}, in which regions of $\ca$, $\cb$ and $\cc$ are shown respectively as green, blue and red sites.

To study the nucleation process that begins in the metastable $\ca$ phase, we must identify clusters of sites that deviate from the $\ca$ phase.  To quantify the cluster properties that correspond to $n$ and $x$ in Eq.~\ref{ours}, we first define a cluster as a contiguous group of $\cb$ (blue) or $\cc$ (red) sites, plus any $\ca$ (green) sites that are completely enclosed by this group of $\cb$ and $\cc$ sites.  The cluster size $n$ is the total number of these $\ca$, $\cb$ and $\cc$ sites, respectively denoted $n^\ca$, $n^\cb$ and $n^\cc$, so that,
\begin{equation}
n=n^\ca+n^\cb+n^\cc.
\end{equation}
We define the core of the cluster as the largest contiguous group of $\cc$ (red) sites that belong to the cluster.  The size of the core $\ncore$ is the number of these $\cc$ sites, plus any $\ca$ or $\cb$ sites that are completely enclosed by the $\cc$ sites of the core.  The cluster composition $x$ is then evaluated using Eq.~\ref{defx}.

Ref.~\onlinecite{James:2019gi} uses approximate definitions for the cluster size and composition, denoted here as $\ntil$ and $\xtil$, and given by,
\begin{eqnarray}
\ntil&=&n^\cb+n^\cc \label{ntil}\\
\xtil&=&n^\cc/\ntil. \label{xtil}
\end{eqnarray}
These approximations were chosen in Ref.~\onlinecite{James:2019gi} for computational efficiency, since both $\ntil$ and $\xtil$ are computed from $n^\cb$ and $n^\cc$ alone, without requiring the identification of the largest $\cc$ region in the cluster, or the relatively rare $\ca$ sites that occur within the cluster.  (Note that in Ref.~\onlinecite{James:2019gi}, $\ntil$ is denoted as ``$n$" and $\xtil$ is denoted as ``$f$".)

In order to maintain consistency with the simulation data of Ref.~\onlinecite{James:2019gi}, 
the new simulations presented here are also conducted using $\ntil$ and $\xtil$ to quantify cluster properties.  Nonetheless, as described in SM Section S1, we derive an approximate transformation that allows us to convert given values of $(\ntil,\xtil)$ 
to corresponding values of $(n,x)$.  This transformation allows us to use simulation results obtained in terms of $(\ntil,\xtil)$ to test predictions expressed in terms of $(n,x)$.  As described in SM, we find that carrying out this transformation is particularly important for accurate estimation of the parameters $S$ and $\xi$.

\section{Simulation results for the free energy surface}

In Ref.~\onlinecite{James:2019gi}, 
umbrella sampling Monte Carlo (MC) simulations were used to generate detailed numerical estimates for the FES describing the TSN process in which the metastable $\ca$ phase converts to the stable $\cc$ phase~\cite{Kumar:1992,Tuckerman:2010,Grossfield:2018}.
This FES
is defined as the free energy of a system of size $N$ in which the {\it largest} cluster occurring in the $\ca$ phase is of size $\ntil$ and composition $\xtil$.  
The transformation given in SM allows us to express this FES in terms of $n$ and $x$.
This transformed FES may be directly compared to $G(n,x)$ as defined in Eq.~\ref{ours} 
for values of $\ntil$ such that the largest cluster in the system is much larger than all other clusters in the system, and
for values of $\xtil$ such that the largest $\cc$-phase region within the largest cluster is much larger than all other $\cc$-phase regions in this cluster.  We find that these conditions are met when $\ntil>500$ and $\xtil>0.05$.  In addition, the FES evaluated in Ref.~\onlinecite{James:2019gi} differs from $G(n,x)$ in Eq.~\ref{ours} by a constant which was not determined in Ref.~\onlinecite{James:2019gi}.  As described in SM Section~\ref{constant}, we have conducted new simulations to determine this constant~\cite{Wolde:1996p3069,Auer:2004db,Lundrigan:2009p5256}.  
After applying these adjustments to the data in Ref.~\onlinecite{James:2019gi}, we present here in 
Fig.~\ref{g3d}(a-c) estimates of $G(n,x)$ 
for three characteristic cases of TSN  
as evaluated directly from the MC simulations described in Ref.~\onlinecite{James:2019gi}.

We also consider the one dimensional (1D) free energy as a function of $n$ alone, which is evaluated from $G(n,x)$ using,
\begin{equation}
\beta {G}_1(n)=-\log \int_0^1 \exp[-\beta {G}(n,x)]\,dx.
\label{g1}
\end{equation}
We define the size $n^*$ of the critical nucleus as the value of $n$ at which $G_1$ is a maximum.
Following Ref.~\onlinecite{James:2019gi} 
we define the average of $x$ at fixed $n$ as,
\begin{equation}
\langle x\rangle=\frac{\int_0^1 x\, \exp[-\beta {G}(n,x)]\,dx}{\int_0^1 \exp[-\beta {G}(n,x)]\,dx}.
\label{xav}
\end{equation}
The fluctuations in $x$, quantified by $\chi=\langle x^2 \rangle - \langle x \rangle^2$, are a maximum at $n=n_c$.  
As explained in Ref.~\onlinecite{James:2019gi},
the significance of $n_c$ is that for $n<n_c$ the most probable state of the nucleus is dominated by the $\cb$ phase, while for $n>n_c$ the most probable nucleus will have a core region of $\cc$ surrounded by $\cb$.  That is, when the nucleus grows to a size greater than $n_c$, a discontinuous phase transition from a pure-$\cb$ nucleus to a core-shell $\cc$-$\cb$ structure becomes possible.
Expressions analogous to Eqs.~\ref{g1} and \ref{xav}, using $(\ntil,\xtil)$ instead of $(n,x)$, are used to define $\ntil^*$ and $\ntil_c$.  
We define the height of the nucleation barrier $G^*$ as the maximum value of $G_1(n)$ or $G_1(\ntil)$, as appropriate.
$G_1(\ntil)$ is shown in Fig.~\ref{gnon}(a) for several values of $H$ as found from MC simulations, and 
Fig.~\ref{gstar} shows the simulation results for $\ntil^*$, $\ntil_c$ and $G^*$ 

\begin{figure}
\centerline{\includegraphics[scale=0.6]{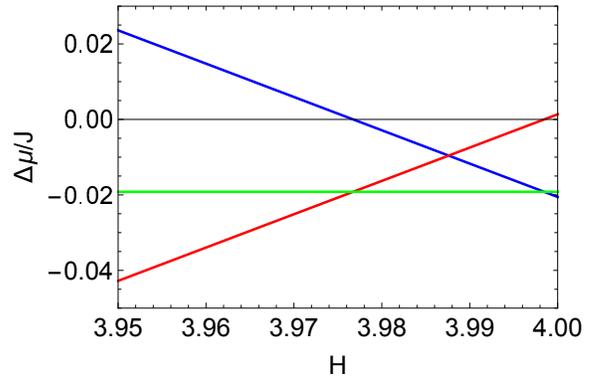}}
\caption{$\Delta \mu_{\ca\cb}$ (blue), $\Delta \mu_{\cb\cc}$ (red)
and $\Delta \mu_{\ca\cc}$ (green)
as a function of $H$ at $H_s=0.01$.}
\label{dmuvH}
\end{figure}

Our goal is to test the degree to which $G(n,x)$ as defined in Eq.~\ref{ours} can predict the results found from MC simulations.  To do so, we require values for the six parameters that occur in Eq.~\ref{ours}:  
$\Delta \mu_{\ca\cb}$, $\Delta \mu_{\cb\cc}$, $\sigma_{\ca\cb}$, $\sigma_{\cb\cc}$, $S$ and $\xi$.  
Ref.~\onlinecite{James:2019gi} provides empirical expressions for 
the chemical potentials of all three phases 
as a function of $H$ and $H_s$ near the triple point.  Fig.~\ref{dmuvH} shows the variation of 
$\Delta \mu_{\ca\cb}$, $\Delta \mu_{\cb\cc}$ and $\Delta \mu_{\ca\cc}$ predicted by these expressions
as a function of $H$ at $H_s=0.01$.
Ref.~\onlinecite{James:2019gi} also finds that $\sigma_{\ca\cb}=\sigma_{\cb\cc}=0.195J$ per unit lattice site of interface, and is independent of $H$ and $H_s$ near the triple point.  Ref.~\onlinecite{James:2019gi} does not provide estimates for $S$ and $\xi$, and so we measure them here, as described in the next section.

\section{Interaction of two interfaces}

\begin{figure}
{\includegraphics[scale=0.4]{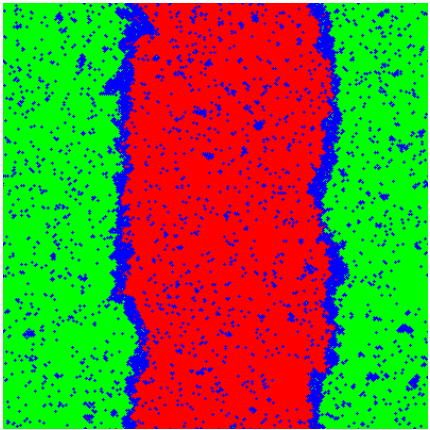}}
\caption{Simulation snapshot of $L=400$ system with planar interfaces separating the $\ca$ phase (green) and $\cc$ phase (red), with a wetting layer of the $\cb$ phase (blue) in between.  This snapshot comes from a run carried out at $H_s=0.01$ and $H=3.985$.  For the large stripe-shaped cluster in the middle of the system, $(\ntil,\xtil)=(80028, 0.871)$ and
$(n,x)=(80322, 0.897)$.
The average width of the wetting layer estimated using Eq.~\ref{dr} is $\Delta r=10.3$.
}
\label{slabs}
\end{figure}

We conduct new simulations of the metamagnet model to estimate the parameters $S$ and $\xi$ that characterize $G_{\rm int}$, the free energy of interaction of the $\ca\cb$ and $\cb\cc$ interfaces.  
The simplest geometry in which to study the interaction of these two interfaces is the ``planar'' case where two flat, parallel interfaces separate semi-infinite regions of the bulk $\ca$ and $\cc$ phases, and where a layer of the $\cb$ phase of width $\Delta r$ lies between the $\ca$ and $\cc$ phases.  An example system configuration having such a planar interface geometry is shown in Fig.~\ref{slabs}.  Since we employ periodic boundary conditions, we simulate a pair of $\ca\cb$-$\cb\cc$ interfaces separated by stripes of the $\ca$ and $\cc$ phases. 

The same umbrella sampling procedure used to study circular clusters in Ref.~\onlinecite{James:2019gi} 
is used here to study the planar interface case.
In the planar case, the largest ``cluster'' in the system is now a rectangular region spanning the periodic boundaries in the vertical direction and consisting of a thick stripe of the $\cc$ (red) phase with thinner wetting layers of the $\cb$ (blue) phase on either side.  
In the planar geometry, the width of each $\cb$ layer, and thus the distance between the $\ca\cb$ and $\cb\cc$ interfaces, is on average,
\begin{equation}
\Delta r=\frac{n(1-x)}{2L}.
\label{dr}
\end{equation}
Umbrella sampling simulations that control both $n$ and $x$ can therefore be used to control $\Delta r$.

Eq.~\ref{ours} models the system free energy when a circular cluster occurs in the $\ca$ phase.  For the planar case, the cluster is a system-spanning rectangular stripe of size $n$ and composition $x$ in a system with periodic boundary conditions.  We denote the free energy of this rectangular cluster as $G_{\parallel}(n,x)$, for which the expression analogous to Eq.~\ref{ours} is, 
\begin{eqnarray}
G_{\parallel}(n,x)&=&n\, \Delta \mu_{\ca\cb} + 2L \sigma_{\ca\cb} \cr
&+&xn\, \Delta \mu_{\cb\cc} + 2L \sigma_{\cb\cc} \cr
&+&2LS\exp[-n(1-x)/2L\xi].
\label{slab}
\end{eqnarray}
We note that Eq.~\ref{slab} assumes that the pair of $\cb$-phase wetting layers (the blue layers in Fig.~\ref{slabs}) are far enough apart so that they do not interact.  Accordingly, we choose $n$ to maximally separate the two wetting layers by $L/2$, and choose $L$ so that 
$L/2$ is much larger than both $\Delta r$ and the observed length scale of the fluctuations of the wetting layers about their mean positions.

\begin{figure}[t]
\centerline{\includegraphics[scale=0.6]{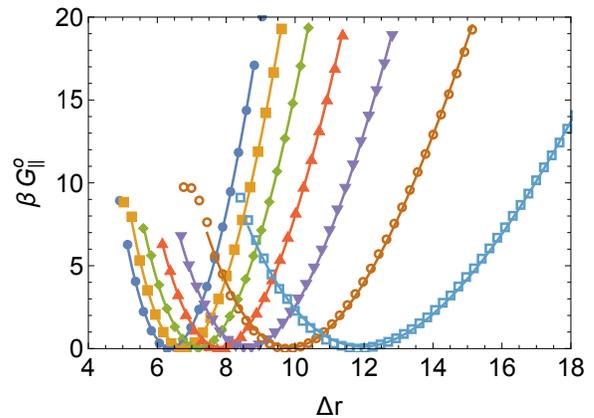}}
\caption{Free energy $G_\parallel^o$ of a system with parallel $\ca\cb$ and $\cb\cc$ interfaces 
separated by a distance
$\Delta r$
for several values of $H$ at $H_s=0.01$.  For curves with minima from left to right, $H=\{3.96, 3.965, 3.97, 3.975, 3.98, 3.985, 3.99\}$.
Solid lines are fits of Eq.~\ref{slab} to the data points, where $S$ and $\xi$ are fit parameters, as described in the text.  Each curve has been shifted by a constant so that $G_\parallel^o=0$ at the minimum.}
\label{gcut}
\end{figure}

\begin{figure}[t]
\centerline{\includegraphics[scale=0.6]{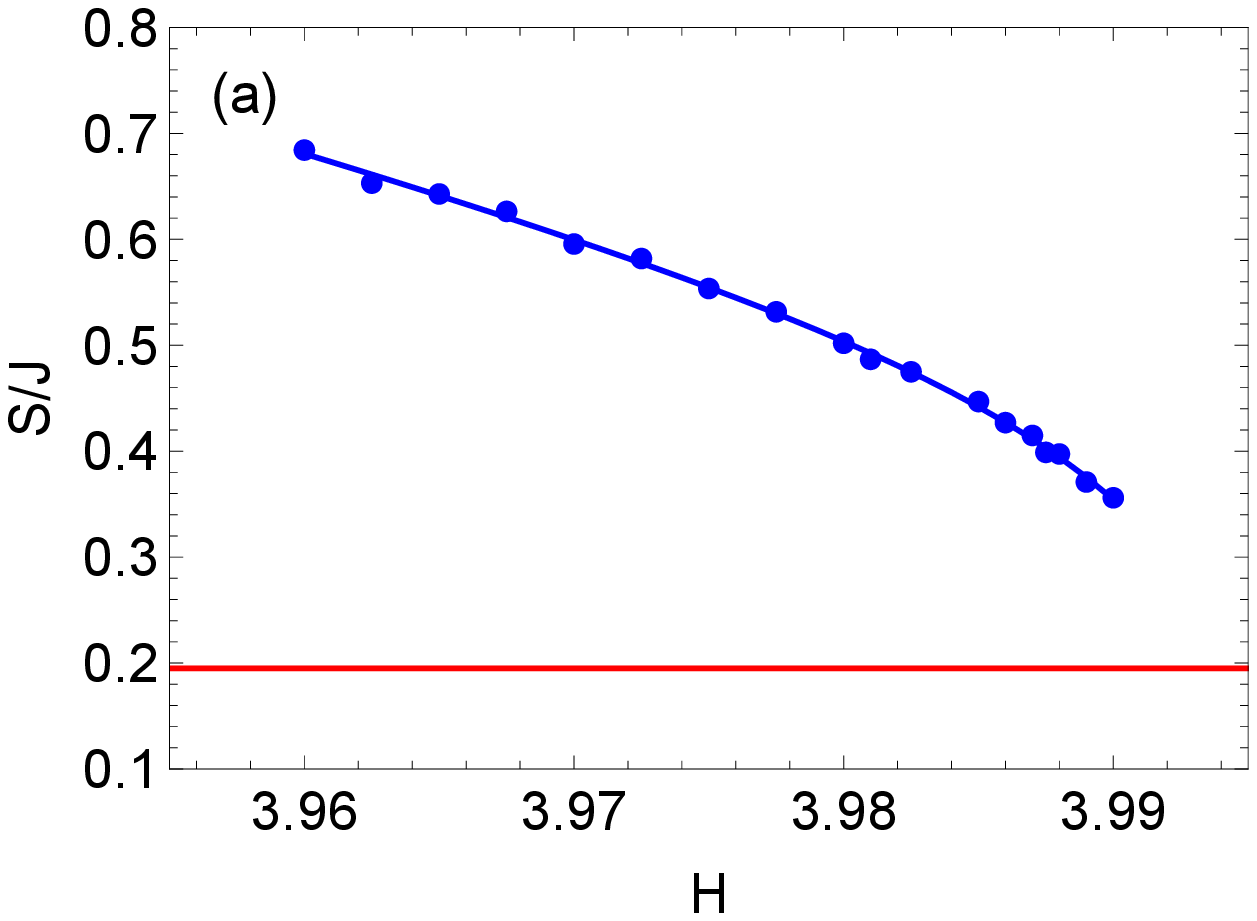}}
\centerline{\includegraphics[scale=0.6]{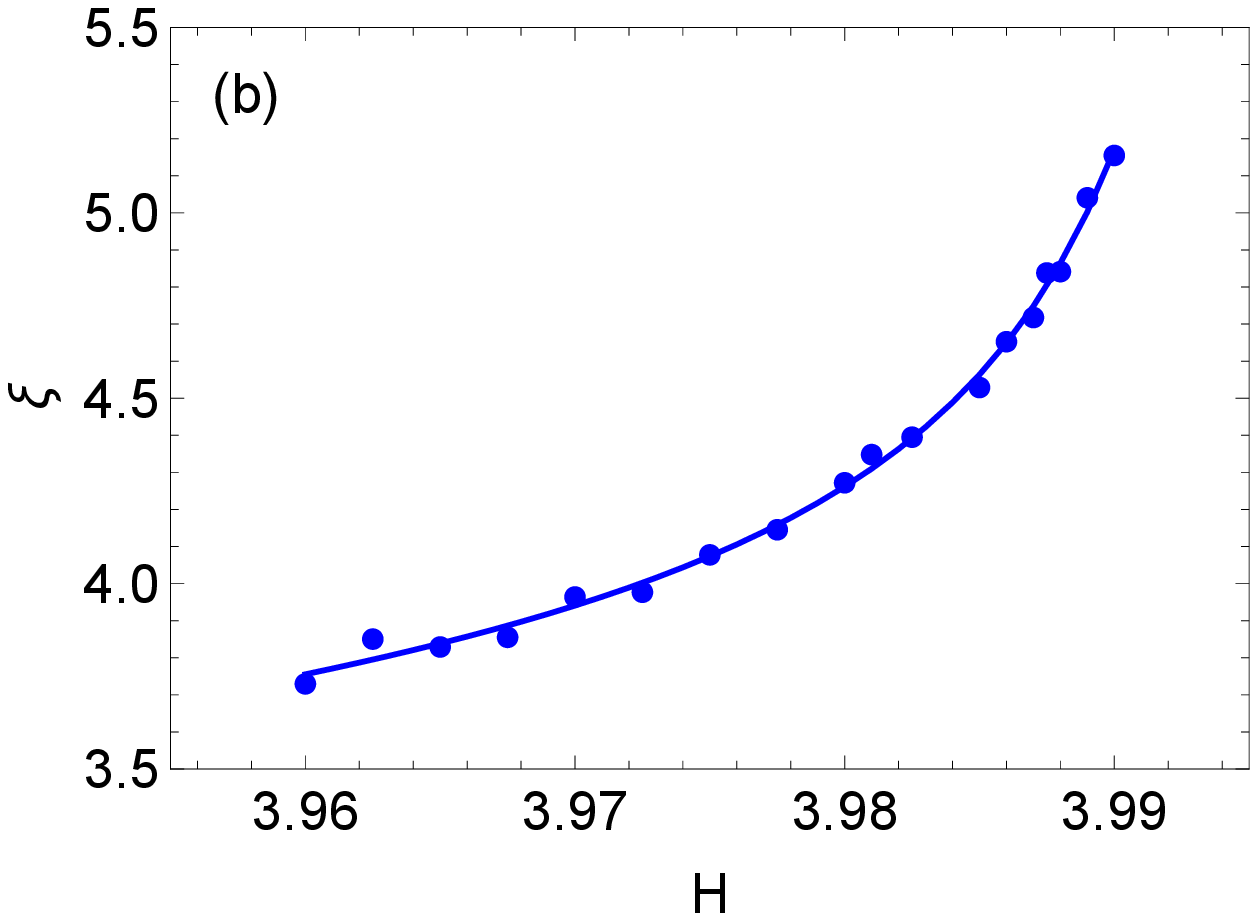}}
\caption{(a) $S$ and (b) $\xi$ as a function of $H$ at $H_s=0.01$,
obtained from fitting Eq.~\ref{slab} to the data for $G_\parallel ^o$ in Fig.~\ref{gcut}.  For comparison, in (a) the horizontal red line shows the value of 
$\sigma_{\ca\cb}=\sigma_{\cb\cc}=0.195J$ per unit lattice site of interface.  In both panels, the solid line is a fit of the empirical expression $a+bH+c/(H-d)$, where $a$, $b$, $c$ and $d$ are fit parameters.
In (a), $(a,b,c,d)=(26.76, -6.572, 0.002005, 4.001)$.
In (b), $(a,b,c,d)=(-19.11, 5.681, -0.01453, 3.999)$.
}
\label{SvH}
\end{figure}

As described in detail in SM Section 3, we carry out umbrella sampling simulations of a system of size $L=400$ (i.e. $N=160\,000$) initialized in the planar geometry~\cite{Binder:2009vp}.  
Fig.~\ref{slabs} is an example configuration resulting from these runs. 
Using a similar simulation protocol as in Ref.~\onlinecite{James:2019gi}, we estimate $G_\parallel(n,x)$, up to an unknown constant $C_\parallel$, by calculating the free energy of a system of size $N$ of the bulk $\ccb$ phase in which the largest cluster in the system is a stripe-shaped cluster of size $\ntil$ and composition $\xtil$.
Using the umbrella sampling method, 
we control $\ntil$ and $\xtil$ so as to vary $\Delta r$ for a system in which 
$\ntil$ remains near the value $\ntil_o=N/2=80\,000$, 
so that approximately half the system is occupied by the cluster.  
As $\xtil$ increases at fixed $\ntil$, the proportion of the cluster occupied by the $\cc$ phase increases but the size of the cluster does not.  The result is that $\Delta r$ decreases as the $\cb$ phase is squeezed out of the region separating $\ca$ from $\cc$.  
From the results of these simulations, and using the transformation from $(\ntil,\xtil)$ to $(n,x)$, we obtain a 1D cut through the FES for 
$G_\parallel (n,x)$ 
along which $\ntil=\ntil_o$, denoted here as $G_\parallel ^o$.
At any point along this 1D cut, we know both $n$ and $x$ and so we can compute $\Delta r$ using Eq.~\ref{dr}.
In Fig.~\ref{gcut}, we plot the result for $G_\parallel^o$ 
as a function of $\Delta r$, obtained at several values of $H$ at fixed $H_s=0.01$.
As expected, $G_\parallel^o$ passes through a minimum corresponding to the equilibrium width of the $\cb$-phase wetting layer.  We observe that the equilibrium width approximately doubles as $H$ varies from $3.96$ to $3.99$.

To estimate $S$ and $\xi$,
we fit $G_\parallel + C_\parallel$
(where $G_\parallel$ is given by Eq.~\ref{slab}) to our simulation data for $G_\parallel ^o$, 
where $S$, $\xi$ and $C_\parallel$ are the fit parameters; see SM Section 3 for details.  The values of $n$ and $x$ used in the fit correspond to their values along the 1D cut that defines $G_\parallel^o$, and the values of $\Delta \mu_{\ca\cb}$, $\Delta \mu_{\cb\cc}$, $\sigma_{\ca\cb}$, $\sigma_{\cb\cc}$ are fixed to those reported in Ref.~\onlinecite{James:2019gi} for the specified values of $H$ and $H_s$.  
The fitted curves for $G_\parallel^o$ are shown in Fig.~\ref{gcut} as solid lines, and the results for $S$ and $\xi$ obtained from these fits are shown in Fig.~\ref{SvH} as a function of $H$ at fixed $H_s=0.01$.  We fit the empirical expression $a+bH+c/(H-d)$, where $a$, $b$, $c$ and $d$ are fit parameters, to our data for both $S$ and $\xi$.  These fitted functions for $S$ and $\xi$, shown in Fig.~\ref{SvH}, allow us to smoothly interpolate the values of $S$ and $\xi$ at arbitrary values of $H$ within the range of our data.

As shown in Fig.~\ref{gcut}, the correspondence between the simulation results and the fitted curves is excellent, confirming that the model of the interface interaction $G_{\rm int}$ given in Eq.~\ref{ours} is accurate and appropriate in this case.  It would be useful for future work to compare this form with results from other approaches that quantify the distance dependence of the interfacial interactions associated with the disjoining pressure~\cite{Bykov:2002hl,Napari:2003js,SHCHEKIN201978}.

\section{Comparison of theory and simulations}

Using the data for $\Delta \mu_{\ca\cb}$, $\Delta \mu_{\cb\cc}$, $\sigma_{\ca\cb}$ and $\sigma_{\cb\cc}$ from Ref.~\onlinecite{James:2019gi}, 
and the results for $S$ and $\xi$ presented here, we have all the parameters required to compute $G(n,x)$ using Eq.~\ref{ours}.  Our results for $G(n,x)$ are shown in the lower panels of Fig.~\ref{g3d} for the same values of $H$ and $H_s$ at which we plot the MC simulation results for $G(n,x)$ in the upper panels.  
Fig.~\ref{g3d} shows that the overall agreement between the prediction of Eq.~\ref{ours} and the MC data is excellent, both in terms of the variation of each FES with $n$ and $x$, and also in terms of how the shape of the FES changes with $H$.  

Fig.~\ref{gnon}(a) 
shows the results for $G_1(\ntil)$ as obtained from Eq.~\ref{g1} when using the prediction of Eq.~\ref{ours}, together with the corresponding results for $G_1(\ntil)$ estimated from simulations.  
In general, the theory tends to underestimate the results for $G_1(\ntil)$ from simulation by several $kT$ and up to $10$~kT under some conditions.  
At the same time, Fig.~\ref{gnon}(a) shows that the characteristic shape of the $G_1(\ntil)$ curves is the same in both the theory and simulations results.  Each $G_1(\ntil)$ curve displays a ``kink" that is a signature of passing through the value $\ntil=\ntil_c$.  The value of $\ntil_c$ increases as $H$ increases and the value of $G^*$ at the maximum of $G_1(\ntil)$ decreases as $H$ increases.  However, the critical size $\ntil^*$ at which the maximum in $G_1(\ntil)$ occurs is not a monotonic function of $H$.  The variation of $n_c$ and $n^*$ as $G_1(n)$ changes with $H$, all computed using Eq.~\ref{ours}, is shown in Fig.~\ref{gnon}(b).  These non-classical features were all noted in Ref.~\onlinecite{James:2019gi}, and the present results show that they also occur in the theory embodied in Eq.~\ref{ours}.

Fig.~\ref{gstar} compares the predictions for $\ntil^*$, $\ntil_c$ and $G^*$
found from $G_1(\ntil)$ using Eq.~\ref{ours} with the values obtained from MC simulations.
Also shown are the values of  $n^*$, $n_c$ and $G^*$ found from $G_1(n)$ using Eq.~\ref{ours}.
While there are systematic differences between theory and simulation for these quantities, the qualitative trends are the same.  In particular, the non-monotonic variation of $n^*$ with $H$ is well reproduced by the theory, as is the accelerating decrease of $G^*$ as $H$ increases in the regime when $n_c>n^*$.  

\begin{figure}[t]
\newcommand\x{0.6}
{\includegraphics[scale=\x]{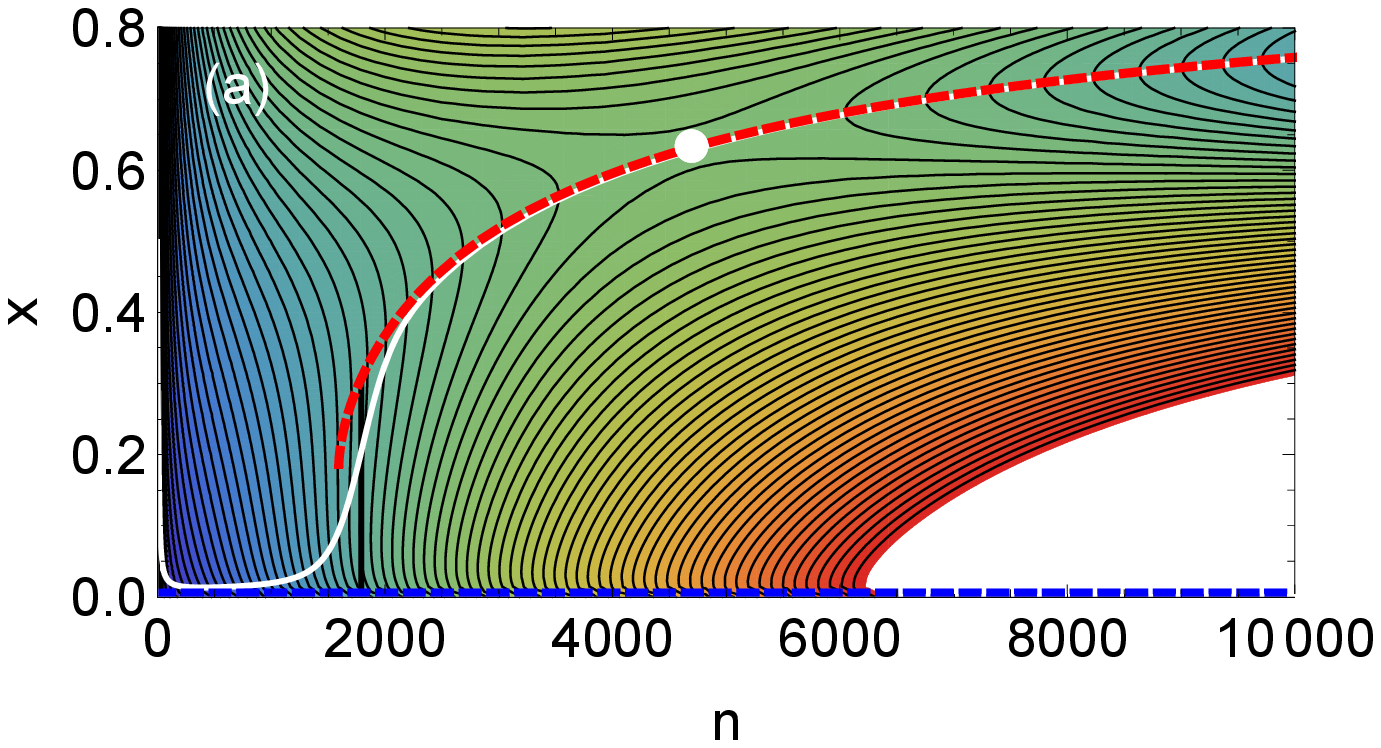}}
{\includegraphics[scale=\x]{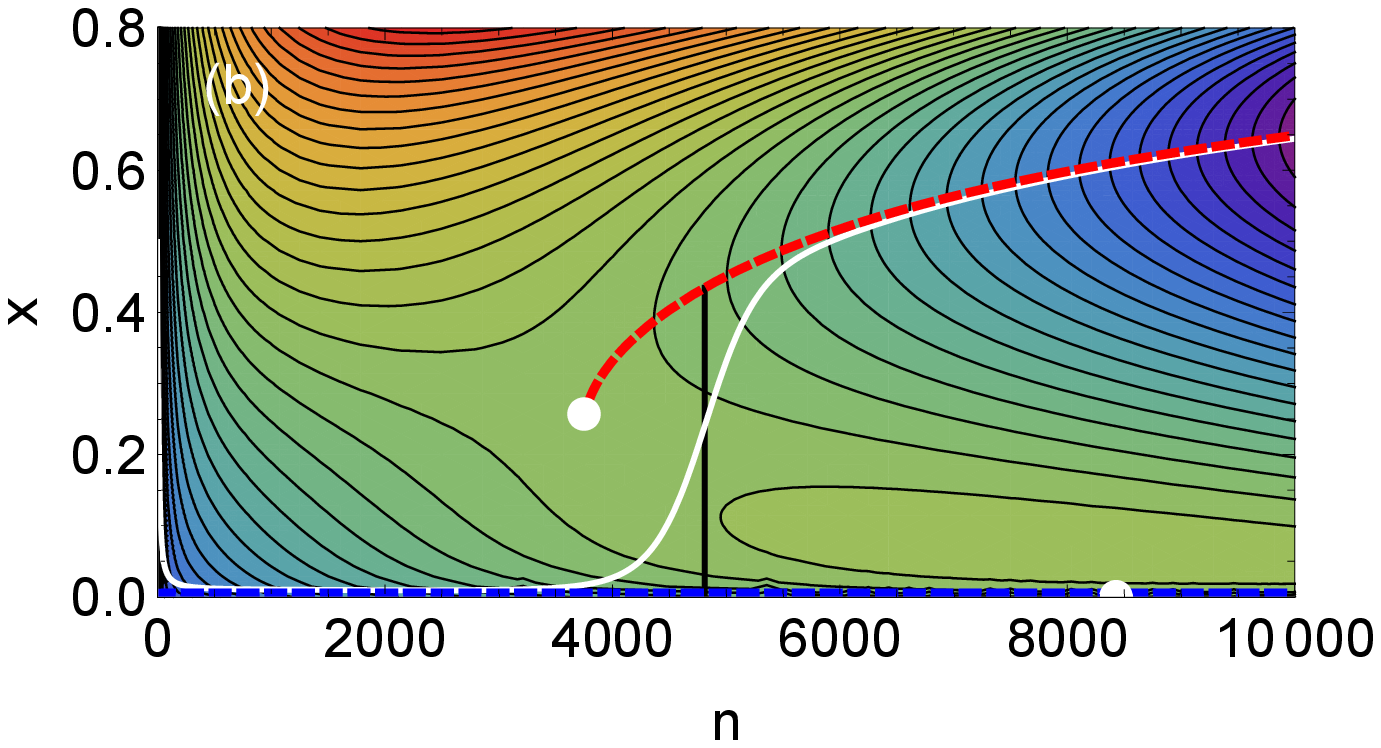}}
{\includegraphics[scale=\x]{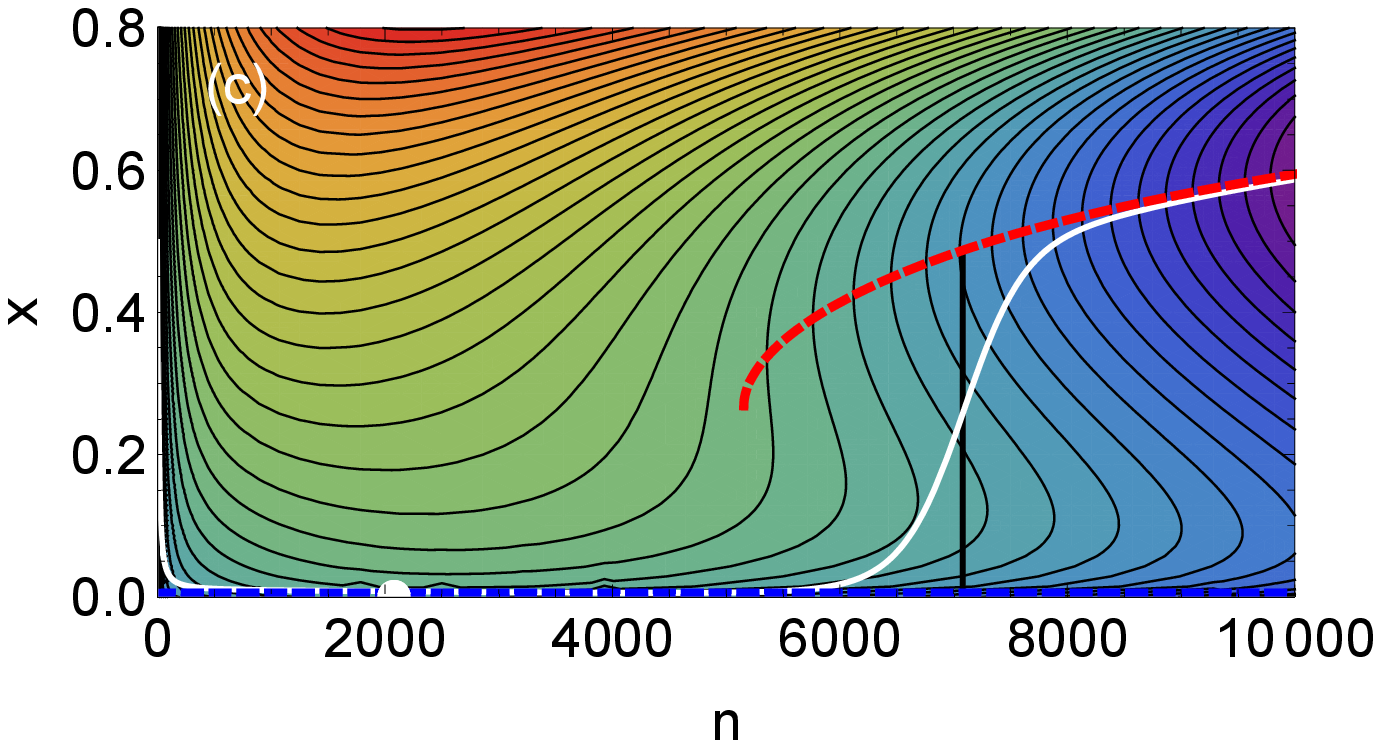}}
\caption{Contour plot of $\beta G(n,x)$ as predicted by Eq.~\ref{ours} for $H_s=0.01$ and for $H=\{3.96, 3.981, 3.985\}$ in (a,b,c) respectively.  In all panels, contours are $2kT$ apart.  White circles are saddle points.  The vertical black line locates $n_c$.  The white curve is $\langle x \rangle$.  The dashed lines locate minima in $G(n,x)$ as a function of $x$ at fixed $n$.  Along the blue dashed line, the cluster is in the $\cb$ phase.  Along the red dashed line, the cluster contains a core of the $\cc$ phase surrounded by a wetting layer of the $\cb$ phase.}
\label{g2d}
\end{figure}

Fig.~\ref{g2d} 
presents three contour plots of the $G(n,x)$ surface obtained using Eq.~\ref{ours}.  These three plots show the FES at the same state points for which analogous plots are shown 
in Fig.~8 of Ref.~\onlinecite{James:2019gi}.  
As shown in Fig.~\ref{g2d}, there are two channels in the FES.  The $\cb$ channel (blue dashed line) begins at $n=0$ and corresponds to the path of a pure $\cb$ phase nucleus growing within the metastable $\ca$ phase.  The $\cc$ channel (red dashed line) always begins at a value of $n>0$ and corresponds to a core-shell nucleus with a $\cc$-phase core surrounded by a shell of the $\cb$ phase.  The most probable small nucleus always appears in the $\cb$ channel and must traverse a ridge in the FES to access the $\cc$ channel.  
The value of $n=n_c$ is indicated by the black vertical line in Fig.~\ref{g2d}.  For $n<n_c$, the pure $\cb$ nucleus is the most stable state of the nucleus at fixed $n$, and the core-shell nucleus is either unstable or metastable.  As $n$ increases through the value of $n_c$, the pure $\cb$ nucleus becomes metastable, and the core-shell nucleus associated with $\cc$ channel becomes the most stable state of the nucleus at fixed $n$.
The transition from the $\cb$ to the $\cc$ channel is therefore probable only when $n>n_c$.
Transition states (saddle points in the FES, indicated by white circles) may occur on the $\cc$ channel [Fig.~\ref{g2d}(a)], 
on the $\cb$ channel [Fig.~\ref{g2d}(c)], or both [Fig.~\ref{g2d}(b)]. 
We plot the variation of $\langle x\rangle$ with $n$ as a white line in Fig.~\ref{g2d}.
This curve represents the average path that would be followed by the system during the nucleation process if the degrees of freedom associated with $x$ are fully equilibrated at each value of $n$.  
For all of the features listed above, the pattern of behavior 
shown here in Fig.~\ref{g2d} 
is also found in Fig.~8 of Ref.~\onlinecite{James:2019gi}.  

\begin{figure}
\newcommand\x{0.45}
{\includegraphics[scale=\x]{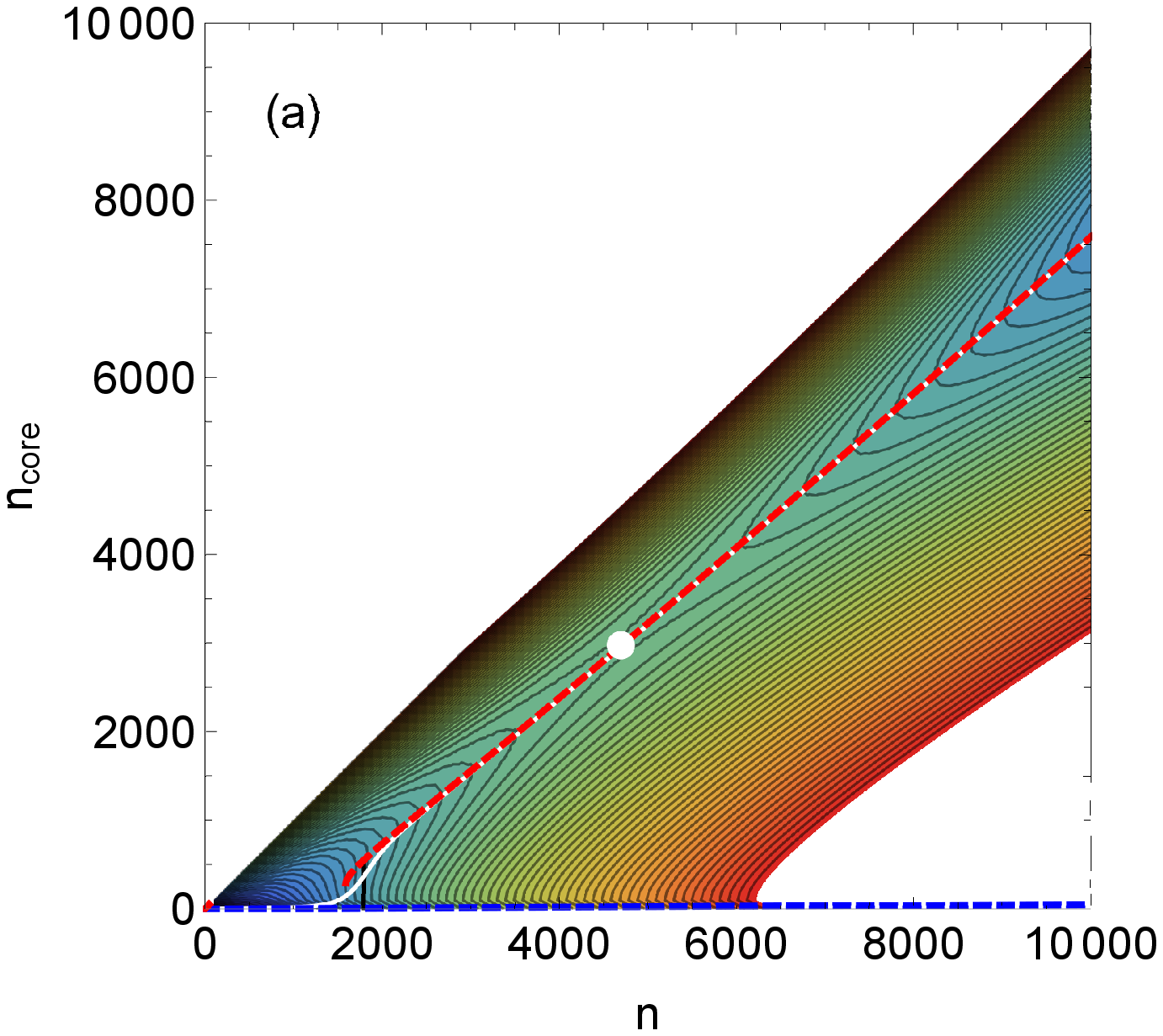}}
{\includegraphics[scale=\x]{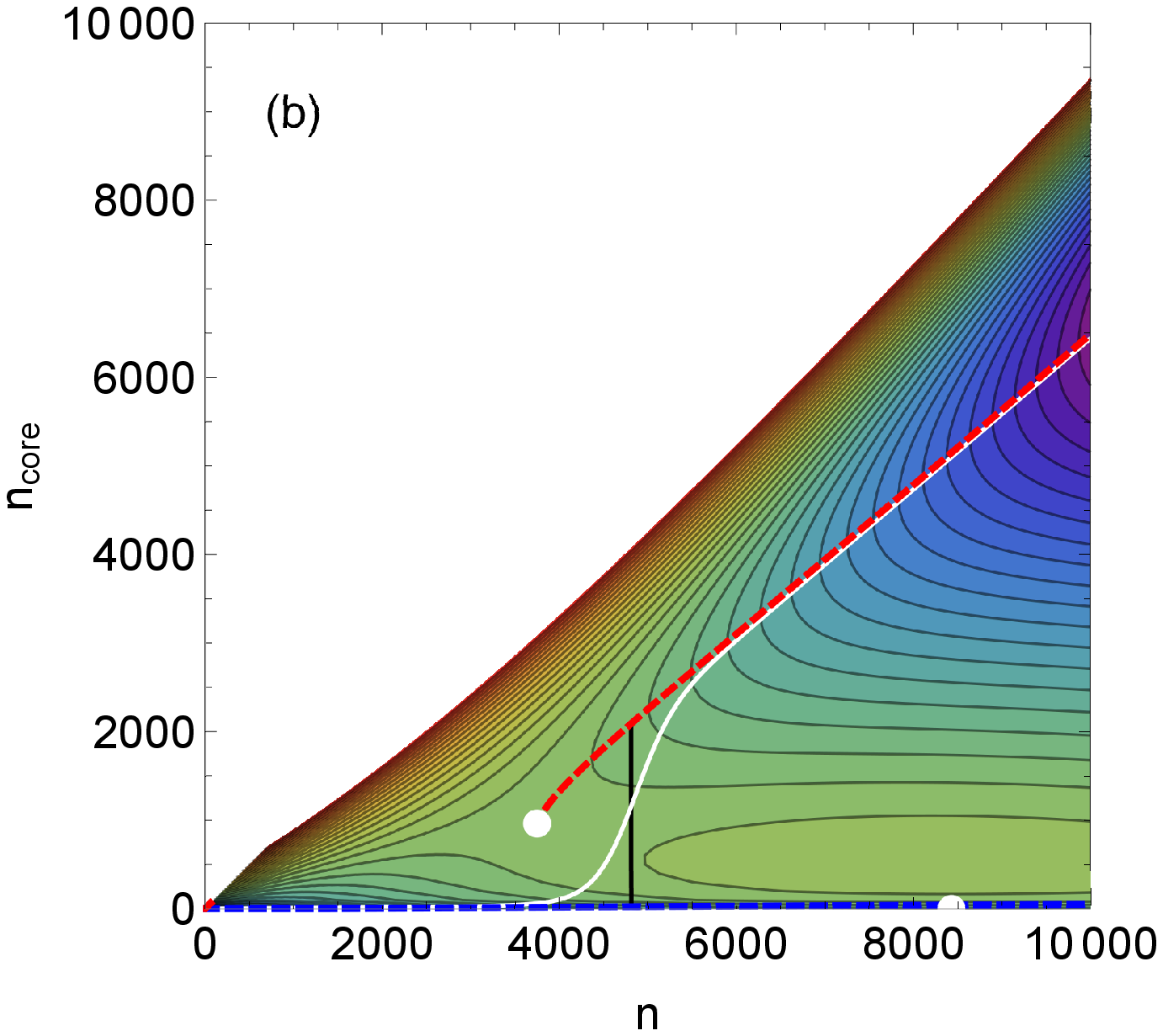}}
{\includegraphics[scale=\x]{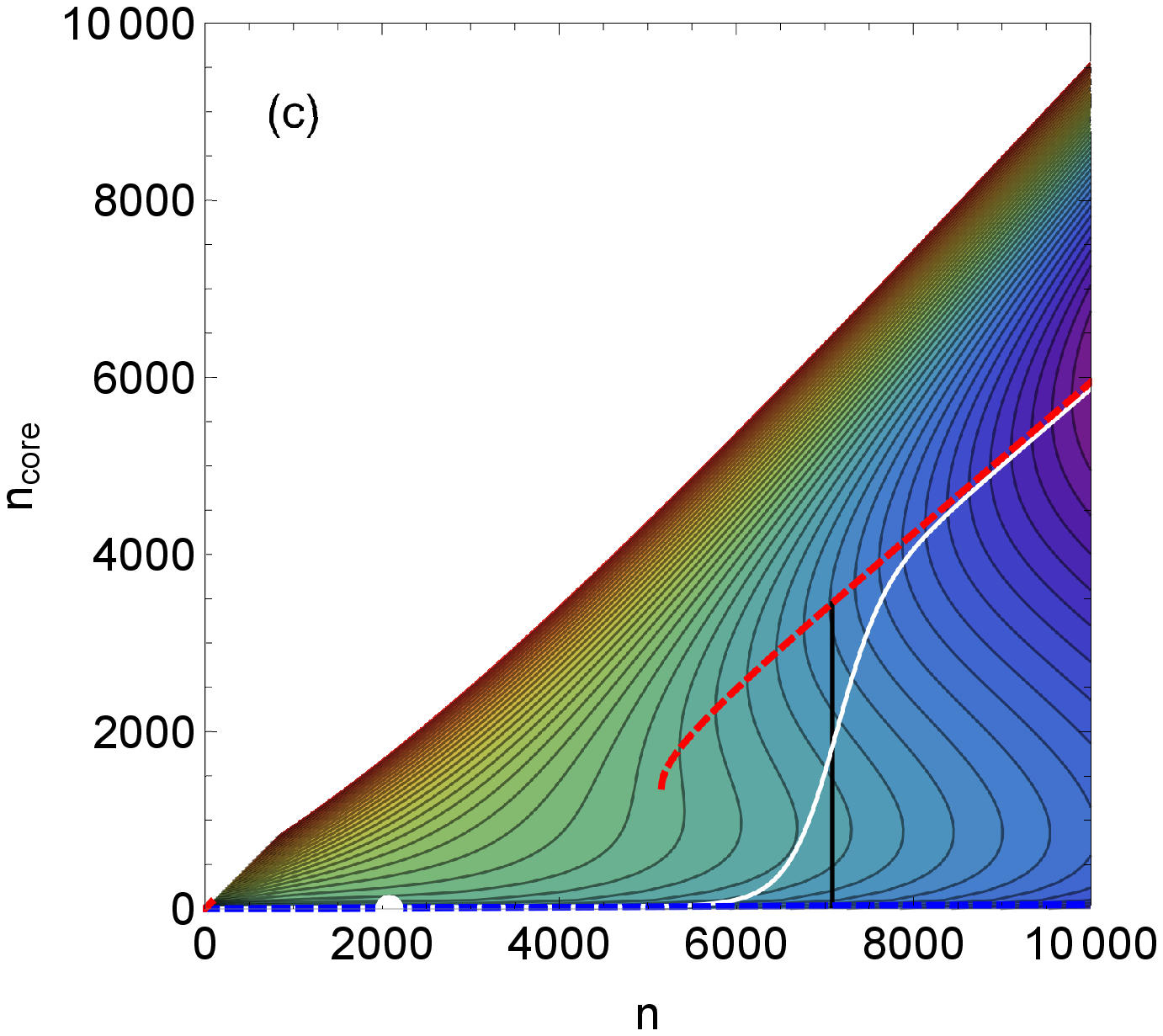}}
\caption{Contour plot of $\beta G(n,\ncore)$ as predicted by Eq.~\ref{ours} for $H_s=0.01$ and for $H=\{3.96, 3.981, 3.985\}$ in (a,b,c) respectively.  In all panels, contours are $2kT$ apart.  Symbols and lines have the same meaning as in Fig.~\ref{g2d}.}
\label{g2dplay}
\end{figure}

\begin{figure}
\newcommand\x{0.75}
{\includegraphics[scale=\x]{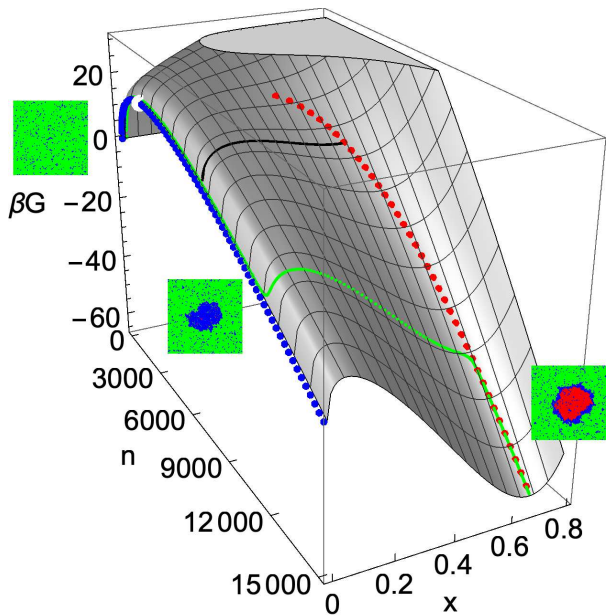}}
\caption{Surface plot of $G(n,x)$ as predicted by Eq.~\ref{ours} for $H_s=0.01$ and for $H=3.985$.
The system at the origin is in the homogeneous metastable $\ca$ phase, as illustrated in the left-most inset image.  Along the channel marked by the blue dotted line, the cluster is in the $\cb$ phase (middle inset) and passes through the saddle point marked by the white circle at the maximum of the blue line.
Along the channel marked by the red dotted line, the cluster is dominated by the $\cc$ phase (right-most inset).
The green dotted line is a schematic TSN pathway, passing first over the saddle point and then over the ridge separating the $\cb$ and $\cc$ channels.  The thick black line locates $n=n_c$, the smallest value of $n$ at which a transition from the $\cb$ to the $\cc$ channel becomes probable.}
\label{thing}
\end{figure}

\section{discussion}

As noted above, there are no adjustable parameters when we compare the FES as obtained from MC simulations and from Eq.~\ref{ours}.  The simulated FES is determined entirely from the microstates of the metamagnet, generated via umbrella sampling runs that explore heterogeneous states in which a localized cluster of size $n$ and composition $x$ occurs in the $\ca$ phase.  The predicted FES obtained from Eq.~\ref{ours} is based on measured
properties of pure homogeneous bulk phases ($\Delta \mu_{\ca\cb}$ and $\Delta \mu_{\cb\cc}$) or properties obtained from systems in which a flat interface separates homogeneous bulk phases ($\sigma_{\ca\cb}$, $\sigma_{\cb\cc}$, $S$ and $\xi$).  Our results thus demonstrate that, similar to CNT for simple (one-step) nucleation, a satisfactory CNT-based theory for the FES of TSN can be constructed using only information on pure bulk phases and flat macroscopic interfaces between bulk phases.

Ref.~\onlinecite{James:2019gi} emphasized the significance of the ridge-crossing process, also observed here, by which the nucleus switches from the $\cb$ channel to the $\cc$ channel.
This process is a discontinuous phase transition that occurs in the finite-sized nucleus as it grows. This is termed a ``fluctuation phase transition" (FPT) in Ref.~\onlinecite{James:2019gi}
because it is a phase transition that occurs in a transient and spatially localized fluctuation, which in the present case is the nucleus.
The FPT is probable only when $n>n_c$.  If $n_c<n^*$ then the FPT occurs as a restructuring of the pre-critical nucleus before it reaches the transition state that represents the exit from the basin of the bulk metastable $\ca$ phase; see Fig.~\ref{g2d}(a).  This case may provide a way to understand non-classical effects observed in pre-critical nuclei that otherwise seem to pass through a transition state typical of simple, one-step nucleation~\cite{Zhou:2019gh}.
Alternatively, when $n_c>n^*$, the nucleus has already passed through the transition state and exited the metastable phase before undergoing the FPT that converts it to a nucleus that contains the stable $\cc$ phase; see Fig.~\ref{g2d}(c).  It is this case that is normally associated with TSN.
In sum, our results show that a FPT is a feature of the FES given by Eq.~\ref{ours} under all conditions studied here and so
may provide a unified explanation of a wide range of non-classical behavior associated with both pre-critical and post-critical nuclei.

We show in Fig.~\ref{g2dplay} 
contour plots of $G$ as defined in Eq.~\ref{ours} plotted in terms of $(n,\ncore)$ rather than in terms of $(n,x)$, for the same state points as in Fig.~\ref{g2d}.  Several previous works have represented the FES of TSN in terms of $(n,\ncore)$, or equivalent variables~\cite{Shchekin:2013ev,Qi:2015iea,Shao:2020ej,Kashchiev:2020iz}, and we provide these plots here to facilitate comparison with these studies.  When comparing the representations of the FES given in Figs.~\ref{g2d} and \ref{g2dplay}, we note that it is easier to resolve the ridge (and the associated FPT) that separates the 
$\cb$ and $\cc$ channels when the FES is plotted in terms of $(n,x)$, especially when $n_c<n^*$.  
Our results show that only the $\cb$ channel of the FES connects to the metastable $\ca$ phase at $n=0$.  That is, the most probable small fluctuations in the metastable phase are those with the lowest surface tension, which here are $\cb$-phase clusters.  Only when the cluster has grown to sizes larger than $n_c$ can the stable $\cc$ phase appear in the nucleus, via the FPT.  This behavior is difficult to resolve when the FES is plotted as in Fig.~\ref{g2dplay}, especially when $n_c<n^*$.
Previous studies of TSN have discussed the possibility that two thermodynamically defined pathways originate from the metastable state at $n=0$ on the FES~\cite{Iwamatsu:2011if,Kashchiev:2020iz}.
The model of the FES given by Eq.~\ref{ours} is not consistent with this picture.

The main characteristics of TSN as described by Eq.~\ref{ours} when $n_c>n^*$ are summarized in Fig.~\ref{thing}.  The FES in Fig.~\ref{thing} 
is the same as that shown as a contour plot in Fig.~\ref{g2d}(c).  As stated above, a single exit pathway (the blue $\cb$ channel) leads out of the metastable state.  The nucleus passes through the transition state (white circle) but remains in the intermediate $\cb$ phase.  The pathway leading to the stable phase (the red $\cc$ channel) can only be reached when $n>n_c$ and via a FPT that carries the nucleus over the ridge in the FES.  Notably, since the transition from the $\cb$ channel to the $\cc$ channel does not pass through a saddle point, knowledge of the FES alone is not sufficient for predicting the size of the nucleus at the FPT.  Rather, the growth dynamics of the nucleus and the relative rates of relaxation of $n$ and $x$ will be controlling factors.  
The FES in Fig.~\ref{thing} thus illustrates how long-lived intermediate-phase nuclei can persist and grow to large size before the stable phase finally appears, a common feature of TSN~\cite{Vekilov:2004jc,Vekilov:2010gm,Sear:2012ji,Lvov:2020kr}.
The green line is an example of such a nucleation trajectory in which the nucleus lingers in the $\cb$ channel well beyond $n_c$ before converting to the $\cc$ channel that leads to the stable phase.  

Our results also confirm that the inclusion of the interaction $G_{\rm int}$ between the $\ca\cb$ and $\cb\cc$ interfaces plays an important role in controlling the shape of the FES for TSN.  A recent study examined the FES formed without including $G_{\rm int}$ and the topography of the surface is distinctly different~\cite{Kashchiev:2020iz}.  In particular, a significant local maximum occurs in the FES that is almost always absent in the FES generated by Eq.~\ref{ours} when $G_{\rm int}$ is included, at least for our system.  It will be interesting for future work to explore the range of FES topographies that result from models of the form of Eq.~\ref{ours} when applied to different systems.

We also note that our study does not address a number of factors that may significantly influence the shape of the FES.  For example, Eq.~\ref{ours} assumes complete wetting of the $\cc$ phase by the $\cb$ phase.  Incomplete wetting changes the geometry of the two-phase nucleus and would require modifications to the form of Eq.~\ref{ours}.   Also, in our system, the volume per monomer does not vary from one phase to another, but this will clearly have an impact on systems in which density is an order parameter, such as crystal formation from a liquid.  We have also not explicitly examined temperature-dependent effects since we have used the thermodynamic fields $H$ and $H_s$ to vary the relative chemical potentials of the three phases involved in our simulation model.  More broadly, the present work does not address the key question of the implications of the FES presented here for the estimation of nucleation rates~\cite{vanMeel:2008hb,Chen:2008dl}.  These are all important avenues for future work.

In summary, our results demonstrate that Iwamatsu's model~\cite{Iwamatsu:2011if} for the FES of TSN works well for the 2D lattice system studied here.  Eq.~\ref{ours} provides a useful qualitative picture of the thermodynamics of TSN, and also yields quantitative predictions that are a satisfactory starting point for estimating the behaviour of a real system.  More generally, our results confirm that significant insights into non-classical nucleation processes can be achieved by an extension of the concepts of traditional CNT to more complex systems.

\section*{Supplemental Material}

The Supplemental Material provides a description of the order parameter transformation from $(\ntil,\xtil)$ to $(n,x)$;  a description of the method used to compute the FES using global order parameters, and how to compare this to the model FES given in Eq.~\ref{ours}; and details related to the estimation of $S$ and $\xi$ from simulations.

\begin{acknowledgments}
We acknowledge the support of the Natural Sciences and Engineering Research Council of Canada (NSERC), Grant Nos. RGPIN-2017-04512 (PHP), RGPIN-2017-05569 (IS), and RGPIN-2019-03970 (RKB).
We also thank ACENET and Compute Canada for support.
\end{acknowledgments}

\section*{Data Availability Statement}

The data that support the findings of this study are available from the corresponding author upon reasonable request.

\bibliography{eaton}



\clearpage

\onecolumngrid

\noindent{\bf \textsf {SUPPLEMENTAL MATERIAL: \\Free energy surface of two-step nucleation}}

\bigskip
\hskip3em{\textsf{D. Eaton$^1$, I. Saika-Voivod$^2$, R.K. Bowles$^3$ and P.H. Poole$^1$}}

\hskip3em{\it $^{1)}$Department of Physics, St. Francis Xavier University, Antigonish, NS, B2G 2W5, Canada}

\hskip3em{\it $^{2)}$Department of Physics and Physical Oceanography, Memorial University of Newfoundland,}

\hskip3em{\it St. John's, Newfoundland A1B 3X7, Canada}

\hskip3em{\it $^{3)}$Department of Chemistry, University of Saskatchewan, Saskatoon, SK, 57N 5C9, Canada}

\bigskip
\hskip3em(Dated: {\today})
\bigskip

\setcounter{equation}{0}
\setcounter{figure}{0}
\setcounter{section}{0}
\setcounter{table}{0}
\setcounter{page}{1}
\makeatletter
\renewcommand{\theequation}{S\arabic{equation}}
\renewcommand{\thefigure}{S\arabic{figure}}
\renewcommand{\thesection}{S\arabic{section}}

\twocolumngrid

\section{Order parameter transformation}
\label{convert}

The order parameters $\ntil$ and $\xtil$ introduced in Ref.~\onlinecite{James:2019gi} 
are related to, but not identical with, the order parameters $n$ and $x$ that appear here in Eq.~\ref{ours}.  There are three reasons for this:  (i) $\ntil$ ignores $\ca$ sites that naturally occur in the $\cb$-dominated shell of the cluster; (ii) $\xtil$ ignores the $\cb$ sites that naturally occur in the $\cc$-dominated core; and (iii) $\xtil$ erroneously includes the $\cc$ sites that naturally occur in the shell when estimating the fractional size of the core.  In an accurate definition of $n$ and $x$, all of the small equilibrium fluctuations that occur in both the core and the shell should be counted as part of each region.  Here we derive the relations that we use to convert values of $(\ntil,\xtil)$ to the corresponding values of $(n,x)$.  

\begin{figure}[b]
{\includegraphics[scale=0.42]{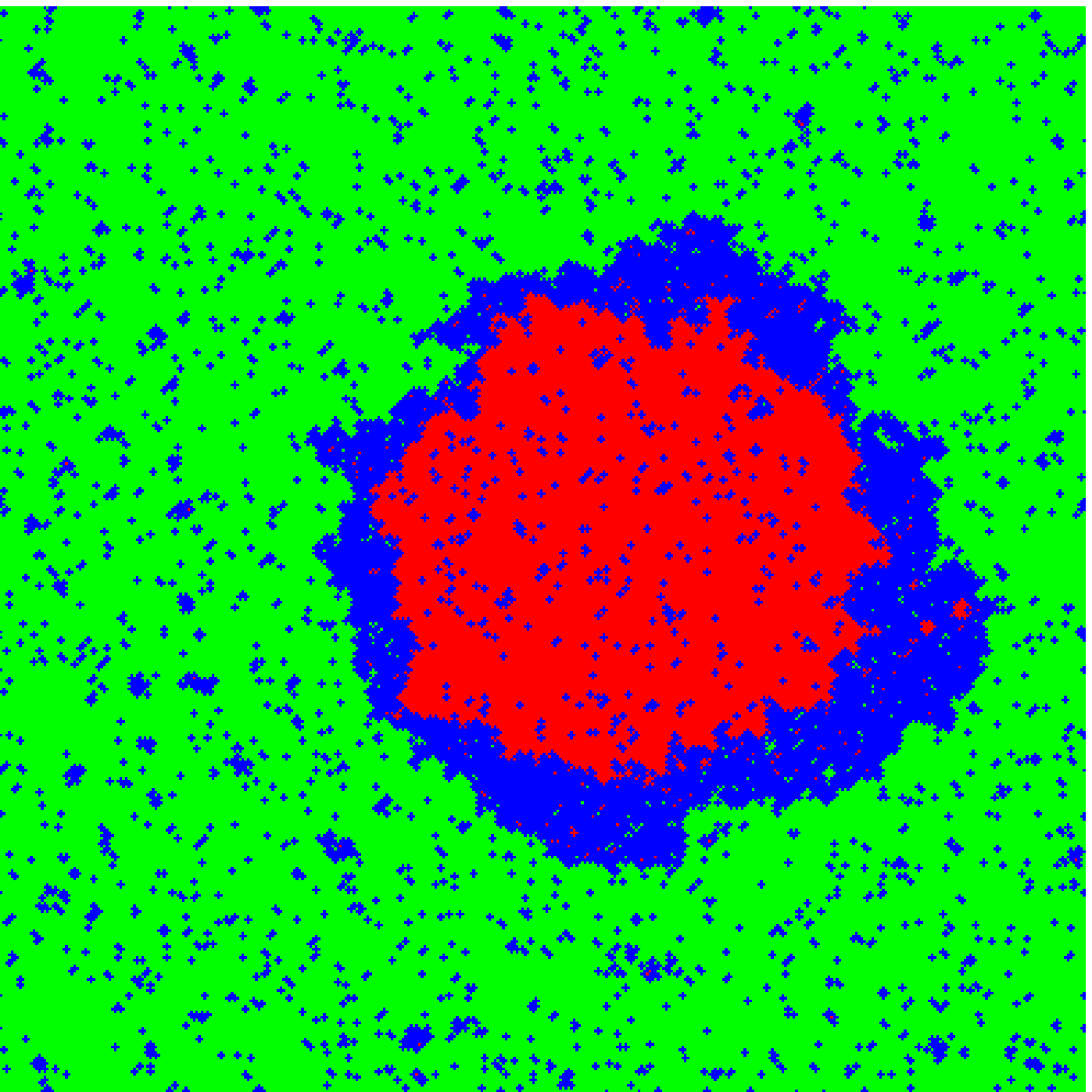}}
\caption{$L=400$ system at $H_s=0.01$ and $H=3.985$ containing a cluster with a core-shell structure with $(\ntil,\xtil)=(40082, 0.598)$ and
$(n,x)=(40670, 0.594)$.}
\label{example}
\end{figure}

Consider our system as a set of sites which are each assigned to the 
$\ca$, $\cb$ or $\cc$ phase.  See for example the system configuration in Fig.~\ref{example}, where $\ca$, $\cb$ and $\cc$ sites are rendered in green, blue and red respectively.  For rendering purposes, the procedure used to decide which sites belong to each phase is described in Ref.~\onlinecite{James:2019gi}.
Let $x^\ca$ be the fraction of sites belonging to the $\ca$ phase, with similar definitions for $x^\cb$ and $x^\cc$ such that,
\begin{eqnarray}
x^\ca+x^\cb+x^\cc=1.
\label{a1}
\end{eqnarray}

We can relate $x^\ca$, $x^\cb$ and $x^\cc$ to the magnetization $m$ and the staggered magnetization $m_s$ of the metamagnet model, defined as,
\begin{eqnarray}
m&=&\frac{1}{N}\sum_{i=1}^N s_i \\
m_s&=&\frac{1}{N}\sum_{i=1}^N \sigma_i s_i.
\label{m}
\end{eqnarray}
We note that regions of $\ca$ or $\cc$ sites are antiferromagnetically ordered with $(m,m_s)=(0,-1)$ or $(m,m_s)=(0,1)$ respectively, and a region of $\cb$ sites is ferromagnetically ordered with $(m,m_s)=(1,0)$.
Therefore, a system made up of regions of $\ca$, $\cb$ and $\cc$ sites will have,
\begin{eqnarray}
 m=x^{\cb}, 
\label{a2}
\end{eqnarray}
 because $m=0$ within the $\ca$ or $\cc$ regions, and $m=1$ within the $\cb$ regions.
Similarly, the system will have 
\begin{eqnarray}
m_s=x^{\cc}-x^{\ca}, 
\label{a3}
\end{eqnarray}
because 
$m_s=1$ within the $\cc$ regions,
$m_s=-1$ within the $\ca$ regions, and
$m_s=0$ within the $\cb$ regions.

We then solve Eqs.~\ref{a1}, \ref{a2} and \ref{a3} for $x^\ca$, $x^\cb$ and $x^\cc$ in terms of $m$ and $m_s$:
\begin{eqnarray}
x^\ca&=&\frac{1}{2}(1-m-m_s) \label{xa} \\[2pt]
x^\cb&=&m \label{xb} \\[2pt]
x^\cc&=&\frac{1}{2}(1-m+m_s) \label{xc}
\end{eqnarray}
That is, the fraction of the system occupied by $\ca$, $\cb$ and $\cc$ sites can be evaluated from the values of $m$ and $m_s$ for the system.

Next consider a cluster of $n$ sites within the system that consists of two sub-regions, a core with $\ncore$ sites and a shell with $\nshell$ sites, such that 
\begin{equation}
n=\ncore+\nshell.
\label{b1}  
\end{equation}
The composition of the cluster is defined as,
\begin{equation}
x=\frac{\ncore}{n}.
\label{b2}
\end{equation}
Every site in the core or the shell is an $\ca$, $\cb$ or $\cc$ site.  Let $\ncoreA$ be the number of $\ca$ sites in the core and $\xcoreA=\ncoreA / \ncore$ be the corresponding fraction of $\ca$ sites in the core.  With similar definitions for both the core and the shell and all three types of site, we have,
\begin{equation}
n=\ncoreA+\ncoreB+\ncoreC+\nshellA+\nshellB+\nshellC
\end{equation}
and,
\begin{eqnarray}
\xcoreA+\xcoreB+\xcoreC&=&1 \\[5pt]
\xshellA+\xshellB+\xshellC&=&1.
\end{eqnarray}

The order parameters $\ntil$ and $\xtil$ (defined in Eqs.~\ref{ntil} and \ref{xtil}) can be expressed in terms of the above quantities by,
\begin{equation}
\ntil=\ncoreB+\ncoreC+\nshellB+\nshellC
\label{nmax}
\end{equation}
and,
\begin{equation}
\xtil=\frac{\ncoreC+\nshellC}{\ntil}
\label{fff}
\end{equation}
For the nucleation process studied here, the core of the cluster that we are concerned with is dominated by the $\cc$ phase and the shell is dominated by the $\cb$ phase.  Further, we find that $\ca$ sites are very rare in the $\cc$-phase core at the conditions we simulate.  Consistent with this observation, we find that $x^\ca<10^{-4}$ in the bulk $\cc$ phase for $H<4$ at $H_s=0.01$.
We therefore set $\ncoreA=0$.  
With this simplification, Eqs.~\ref{nmax} and \ref{fff} can be rewritten as,
\begin{eqnarray}
\ntil&=&\ncore + (\xshellB+\xshellC)\nshell  \\[5pt]
\ntil \, \xtil&=&\xcoreC \ncore+\xshellC \nshell
\end{eqnarray}
Solving the above equations for $\ncore$ and $\nshell$, and then using the results in Eqs.~\ref{b1} and \ref{b2}, leads to the following expressions for $n$ and $x$ in terms of $\ntil$ and $\xtil$:
\begin{eqnarray}
n&=&\ntil \, \frac {\xcoreC+\xtil(\xshellB-1)+(\xtil-1)\xshellC} {\xcoreC(\xshellB+\xshellC)-\xshellC} \label{ntrue} \\[5pt]
x&=&\frac{\xtil \, \xshellB+(\xtil-1)\xshellC}{\xcoreC+\xtil(\xshellB-1)+(\xtil-1)\xshellC} \label{xxx}
\end{eqnarray}

For a given value of $H$ and $H_s$, the values of $\xcoreC$, $\xshellB$ and $\xshellC$ may be found using Eqs.~\ref{xb} and \ref{xc} in the following way.  
The core is dominated by the $\cc$ phase and so if we know $m$ and $m_s$ for the homogeneous bulk $\cc$ phase under the same conditions, which we denote $m^\cc$ and $m^\cc_s$,
then we can find $\xcoreC$ using Eq.~\ref{xc}.  
Similarly, the shell is dominated by the $\cb$ phase and so if we know $m$ and $m_s$ for the homogeneous bulk $\cb$ phase, 
which we denote $m^\cb$ and $m^\cb_s$,
we can find $\xshellB$ using Eq.~\ref{xb} 
and $\xshellC$ using Eq.~\ref{xc}.

\begin{figure}
\centerline{\includegraphics[scale=0.6]{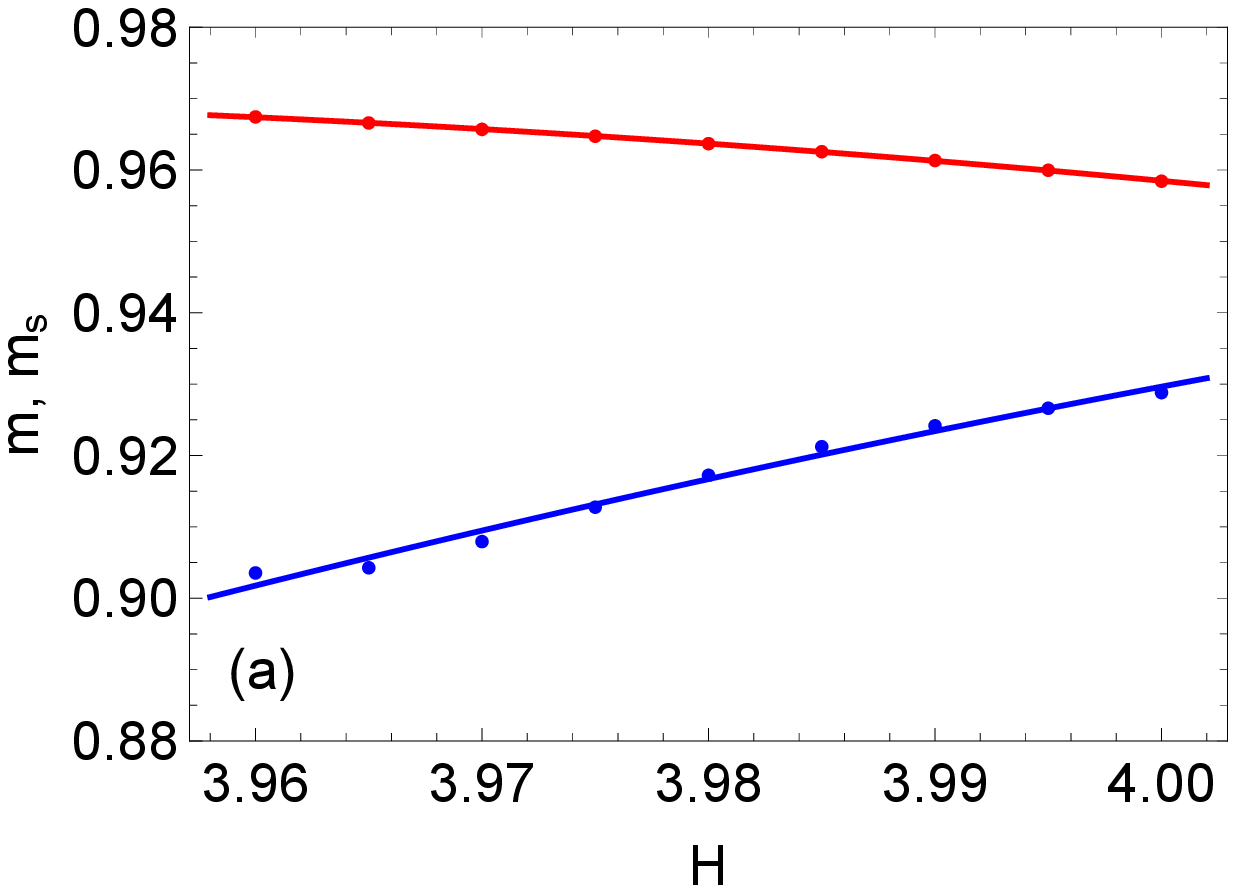}}
\centerline{\includegraphics[scale=0.6]{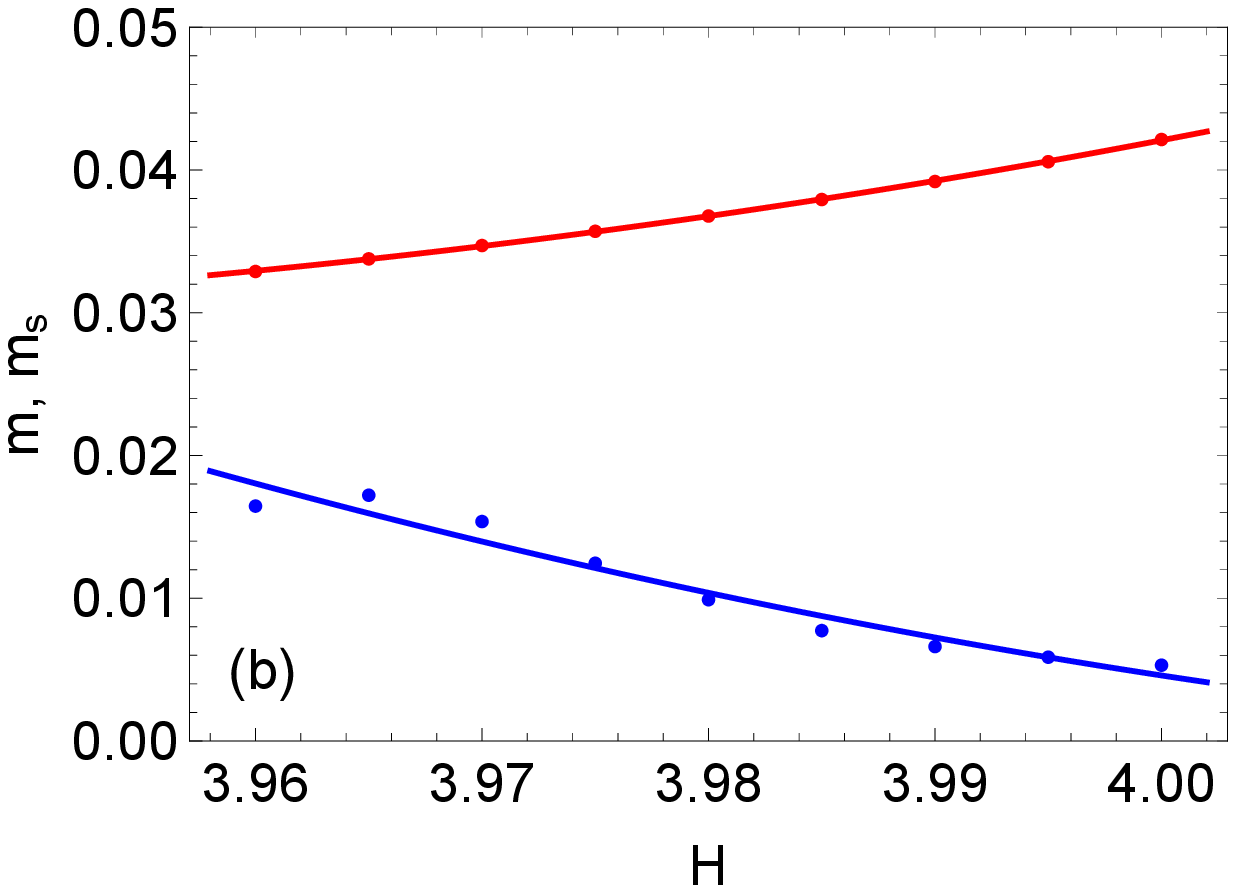}}
\caption{Plots of (a) $m^\cb$ (blue) and $m_s^\cc$ (red), and (b) $m^\cc$ (red) and $m_s^\cb$ (blue), versus $H$ at $H_s=0.01$.  Data points are found using Eqs.~\ref{intm} and \ref{intms}.  Lines are fits of a quadratic polynomial in $H$.}
\label{mvH}
\end{figure}

We evaluate $m$ and $m_s$ for the bulk $\cb$ and $\cc$ phases using the free energy surface $G(m_s,m)$ presented in the Supplemental Material for Ref.~\onlinecite{James:2019gi}.  As described in Ref.~\onlinecite{James:2019gi}, $G(m_s,m)$ can be reweighted to any value of $H$ and $H_s$ near the triple point.  The thermodynamic average of $m$ or $m_s$ for a given phase can then be found
by integration over $G(m_s,m)$: 
\begin{eqnarray}
m&=&\frac{\int_{m_s^*}^{1}dm_s\int_{m^*}^{1}dm\,  m \exp[-\beta G(m_s,m)]}
{\int_{m_s^*}^{1}dm_s\int_{m^*}^{1}dm\,  \exp[-\beta G(m_s,m)]} \label{intm} \\
m_s&=&\frac{\int_{m_s^*}^{1}dm_s\int_{m^*}^{1}dm\,  m_s \exp[-\beta G(m_s,m)]}
{\int_{m_s^*}^{1}dm_s\int_{m^*}^{1}dm\,  \exp[-\beta G(m_s,m)]} \label{intms}
\end{eqnarray}
The lower limits of integration $m_s^*$ and $m^*$ are chosen to restrict the integration to the basin in $G(m_s,m)$ corresponding to the desired phase.  
To find $m^\cc$ and $m_s^\cc$, we use $m_s^*=0.9$ and $m^*=0$.
To find $m^\cb$ and $m_s^\cb$, we use $m_s^*=-1$, but we must take care with the choice of $m^*$ because the bulk $\cb$ phase is approaching its limit of stability as $H$ decreases in the range $3.96<H<4$ at $H_s=0.01$.  The basin in $G(m_s,m)$ corresponding to the $\cb$ phase is shrinking rapidly in this range, and we therefore adjust $m^*$ for each choice of $H$ to ensure that the integration over $G(m_s,m)$ includes only those values of $m$ within the $\cb$ basin.

Fig.~\ref{mvH} shows how
$m^\cb$, $m_s^\cb$, $m^\cc$ and $m_s^\cc$ 
vary with $H$ at $H_s=0.01$, when calculated as described above.  As a check, we have confirmed many of these data points from direct simulations of the bulk $\cb$ and $\cc$ phases.  We note that we are not able to use direct simulations to obtain values of $m$ and $m_s$ for the bulk $\cb$ phase for $H<3.98$ at $H_s=0.01$ because bulk $\cb$ rapidly transforms to the $\cc$ phase under these conditions.  This limitation is the reason we have used Eqs.~\ref{intm} and \ref{intms} to estimate $m^\cb$ and $m_s^\cb$ when approaching the limit of stability of the bulk $\cb$ phase.

The solid lines in Fig.~\ref{mvH} are
fits of a quadratic polynomial in $H$ to each data set.  
These fitting functions allow us to implement the order parameter transformation in Eqs.~\ref{ntrue} and \ref{xxx} for arbitrary values of $H$ in the range $3.96<H<4$ at $H_s=0.01$.

Having defined the transformation from 
$(\ntil,\xtil)$ to 
$(n,x)$, 
we assess the difference it makes to our results.  
The values of $(n,x)$ themselves do not differ greatly from $(\ntil,\xtil)$.  The difference between $n$ and $\ntil$ at fixed $\xtil$ is never more than $4\%$, and the difference between $x$ and $\xtil$ is never more than $0.05$.  The values of the bulk and surface terms in Eq.~\ref{ours} are therefore not greatly affected by the transformation.  However, our estimates of $S$ and $\xi$ obtained by fitting simulation data to Eq.~\ref{slab} depend on an estimate of the interface separation $\Delta r$, which depends on $n$ and $x$ as described in Eq.~\ref{dr}.  The range of $x$ over which we carry out the fit to find $S$ and $\xi$ (see SM Section S3) corresponds to values of $\Delta r<20$ as shown in Fig.~\ref{gcut}.  In this range, we find that the values of $\Delta r$ 
found using $(\ntil,\xtil)$ versus $(n,x)$ differ by up to $40\%$.  As a result, the estimates obtained for $S$ and $\xi$ differ significantly depending on whether or not the transformation from 
$(\ntil,\xtil)$ to $(n,x)$ is used.

Furthermore, we note that the estimates for the chemical potential differences and surface tensions used in Eq.~\ref{ours} are based on calculations that use the bulk order parameters $m$ and $m_s$, as described in Ref.~\onlinecite{James:2019gi}.  These quantities thus incorporate the influence of the fluctuations that are neglected in the definitions of $\ntil$ and $\xtil$.  The estimates of $S$ and $\xi$ obtained using $(n,x)$ therefore correspond better with the other physical parameters used in Eq.~\ref{ours} than the estimates for $S$ and $\xi$ obtained using $(\ntil,\xtil)$.  For these reasons, in the main paper we use the results for $S$ and $\xi$ obtained using the order parameters $(n,x)$, and we present our results in terms of $(n,x)$ whenever possible.


\section{Comparing free energy surfaces from MC simulations and theory}
\label{constant}

Both in Ref.~\onlinecite{James:2019gi} 
and in the present work, the FES evaluated from MC simulations is obtained in terms of global (i.e. system-level) order parameters that correspond to the size ($\ntil$) and composition ($\xtil$) of the {\it largest} cluster in the system. 
We denote the FES that we compute directly from MC simulations as $\cg(\ntil,\xtil)$.
That is, to quantify the thermodynamic properties of the clusters that occur in the $\ca$ phase, we evaluate $\cg(\ntil,\xtil)$, 
the FES of a system of size $N$ of the bulk $\ccb$ phase in which the largest cluster in the system has size $\ntil$ and composition $\xtil$~\cite{Duff:2009p6360}.

We obtain $\cg(\ntil,\xtil)$ from umbrella sampling MC simulations at fixed $(N,H_s,H,T)$~\cite{Tuckerman:2010}.
$\cg(\ntil,\xtil)$ is computed using,
\begin{equation}
\beta \cg(\ntil,\xtil)=-\log [P(\ntil,\xtil)] + C_o,
\label{gnf}
\end{equation}
where $P(\ntil,\xtil)$ 
is proportional to the probability to observe a system microstate in which the largest cluster is of size $\ntil$ and composition $\xtil$. 
The value of the constant $C_o$
is chosen so that $\cg=0$ at the local minimum of $\cg(\ntil,\xtil)$ that occurs close to the origin at $(\ntil,\xtil)=(0,0)$.

We estimate
$P(\ntil,\xtil)$ from 2D umbrella sampling simulations using a biasing potential that depends on both $\ntil$ and $\xtil$,
\begin{equation}
U_B=\kappa_n(\ntil-\ntil')^2 + \kappa_x(\xtil-\xtil')^2,
\label{us}
\end{equation}
where $\ntil'$ and $\xtil'$ are target values of $\ntil$ and $\xtil$ to be sampled in a given umbrella sampling simulation, and 
$\kappa_n$ and $\kappa_x$ control the range of sampling around $\ntil'$ and $\xtil'$.  
Results from multiple umbrella sampling runs conducted at fixed $(N,H_s,H,T)$
are combined using the weighted histogram analysis method (WHAM) to estimate the full $\cg(\ntil,\xtil)$ FES at a given state point~\cite{Kumar:1992,Tuckerman:2010,Grossfield:2018}.

\begin{figure}
{\includegraphics[scale=0.6]{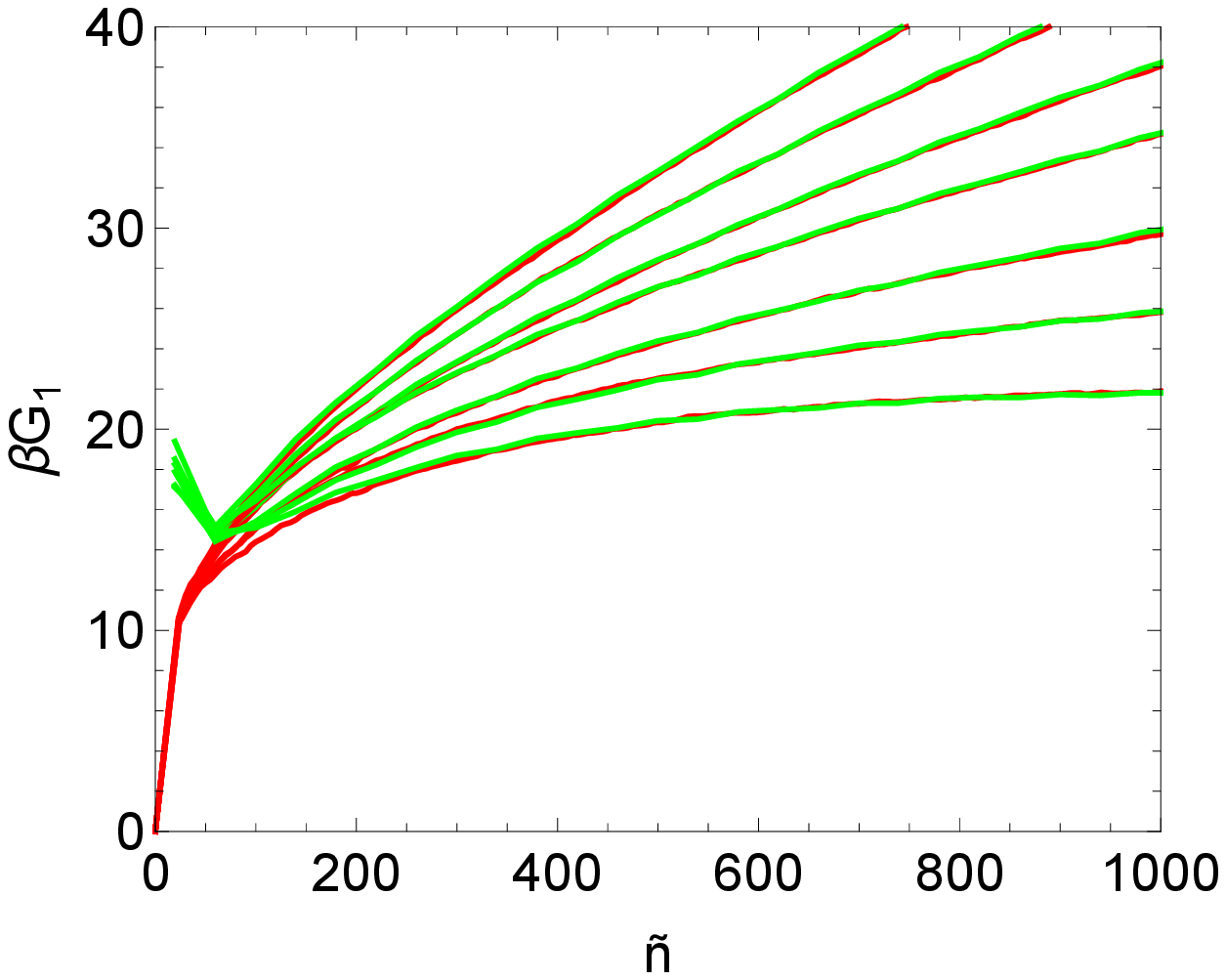}}
\caption{$\beta \cg_1(\ntil)+\beta C$ (green) and $-\log[{\cal N}(\ntil)/N]$ (red) for $H_s=0.01$ and $H=3.96$ to $3.99$ in steps of $0.005$ from top to bottom.}
\label{gnonc}
\end{figure}

In order for $\cg(\ntil,\xtil)$ to correspond to $G(\ntil,\xtil)$ as defined in Eq.~\ref{ours}, $\ntil$ must be large enough so that the largest cluster in the system is much larger than any other cluster in the system, and $\xtil$ must be large enough so that the largest region of the $\cc$ phase within the largest cluster is itself much larger than any other $\cc$ phase region within this cluster.  We find that these conditions are satisfied when $\ntil>500$ and $\xtil>0.05$.  Under these conditions the relationship between $\cg(\ntil,\xtil)$ and $G(\ntil,\xtil)$ is simply,
\begin{equation}
G(\ntil,\xtil)=\cg(\ntil,\xtil)+C,
\label{gg}
\end{equation}
where the constant $C$ depends on the arbitrary choice for the zero of $\cg$ that arises when doing a particular simulation.

We can find the value of $C$ by considering the 1D free energy function defined in Eq.~\ref{g1}.  When considering $\cg(\ntil,\xtil)$ and $G(\ntil,\xtil)$, 
the corresponding 1D free energies are defined respectively as,
\begin{equation}
\beta {\cg}_1(\ntil)=-\log \int_0^1 \exp[-\beta {\cg}(\ntil,\xtil)]\,d\xtil,
\label{g100}
\end{equation}
and
\begin{equation}
\beta {G}_1(\ntil)=-\log \int_0^1 \exp[-\beta {G}(\ntil,\xtil)]\,d\xtil,
\end{equation}
from which it follows that,
\begin{equation}
\beta G_1(\ntil)=\beta \cg_1(\ntil)+\beta C,
\label{gone}
\end{equation}
where $C$ has the same value as in Eq.~\ref{gg}.  Furthermore, $G_1(\ntil)$ can be measured directly in simulations using,
\begin{equation}
\beta G_1(\ntil)=-\log\frac{{\cal N}(\ntil)}{N},
\label{non}
\end{equation}
where ${\cal N}(\ntil)$ is the average number of clusters of size $\ntil$ in a system of size $N$~\cite{Wolde:1996p3069,Auer:2004db,Lundrigan:2009p5256}.  We can therefore estimate $C$ from,
\begin{equation}
\beta C=-\log\frac{{\cal N}(\ntil)}{N} - \beta \cg_1(\ntil),
\label{shift}
\end{equation}
so long as we choose a value of $\ntil>500$ at which ${\cal N}(\ntil)$ may also be reliably evaluated.

To measure ${\cal N}(\ntil)/N$, 
we conduct 1D umbrella sampling simulations with respect to $\ntil$ only, using the same procedure described in detail in Section~S7 of the SM of Ref.~\onlinecite{James:2019gi}.  We conduct these simulations for a system of size $L=200$ at $H_s=0.01$ for $H=3.96$ to $3.99$ in steps of $0.005$.  We obtain $\cg_1(\ntil)$ 
for the same range of $H$ from the data for $\cg(\ntil,\xtil)$ 
using Eq.~\ref{g100}.  Our results for 
${\cal N}(\ntil)/N$ and $\cg_1(\ntil)$ 
are shown in Fig.~\ref{gnonc}, where $\cg_1(\ntil)$ has been shifted by the value of $C$ found using Eq.~\ref{shift} with the choice $\ntil=600$.  Fig.~\ref{gnonc} confirms that $-\log[{\cal N}(\ntil)/N]$ and $\beta \cg_1(\ntil)+\beta C$ coincide for $\ntil>500$.  We also note that ${\cal N}(\ntil)$ may be reliably evaluated using 1D umbrella sampling for $\ntil<1000$ because $\cg(\ntil,\xtil)$ exhibits only one minimum (near $\xtil=0$) with respect to $\xtil$ in this range.  The sampling of the ${\cal N}(\ntil)$ distribution in this range is thus not complicated by the presence of the other minimum that appears at larger $\ntil$.  We find that for all $H$ studied, the value of $C$ is approximately constant with $\beta C=6.7\pm 0.4$.

Having evaluated $C$, we find $G(\ntil,\xtil)$ from $\cg(\ntil,\xtil)$ using Eq.~\ref{gg}.  We then convert this estimate of $G(\ntil,\xtil)$ to an estimate for $G(n,x)$ using the transformation described in Section~\ref{convert}.
We are thus able to make a direct comparison, shown in Fig.~\ref{g3d}, of $G(n,x)$ as found from MC simulations with the prediction given by Eq.~\ref{ours}.  

The simulation results for $G_1(\ntil)$ shown in Fig.~\ref{gnon} are formed by splicing together the curves shown in Fig.~\ref{gnonc}
for ${\cal N}(\ntil)/N$ for $\ntil<600$ 
with our results for $\cg_1(\ntil)+C$ for $\ntil>600$.

\section{umbrella sampling simulations of a planar interface}
\label{2d}

To estimate the parameters $S$ and $\xi$ that appear in Eq.~\ref{ours}, we conduct 2D umbrella sampling simulations of the kind described in Section~\ref{constant} to find $\cg(\ntil,\xtil)$ for a system in which the $\ca$ and $\cc$ phases are separated by a planar interface containing a thin wetting layer of the $\cb$ phase.

We choose $L=400$, $\kappa_n=0.0005J$ and 
$\kappa_x=500J$.  
For each choice of $(N,H_s,H,T)$
we conduct 60 simulations for 
$\ntil' \in \{79900,80000,80100\}$, and for 
$\xtil'=i/20$ where the integer $i\in \{0,1,2,\dots,19\}$.  To study a system with planar interfaces, each run is initiated from a perfect $\ca$ configuration, into which a thick vertical stripe of the perfect $\cc$ phase has been inserted.  A thinner vertical stripe of the perfect $\cb$ phase is then inserted at the two $\ca\cc$ interfaces.  In any given run, the number of $\cb$ and $\cc$ sites inserted is chosen so that $\ntil$ is closest to $\ntil'$ and 
so that the proportion of $\cb$ and $\cc$ sites gives a value of $\xtil$ closest to $\xtil'$.  Our choice of values for $L$ and $\ntil'$ generate system configurations in which the $\cb$-phase wetting layers are separated by approximately $L/2$.  The thickness of each wetting layer is controlled by the choice $\xtil'$.

This system is equilibrated for $5\times 10^4$~MCS (Monte Carlo steps), and then the time series of $\ntil$ and $\xtil$ 
is recorded every 100~MCS for $10^6$~MCS.
We sample configurations using Metropolis single-spin-flip MC dynamics~\cite{Binder:2009vp}.  
Trial configurations are accepted or rejected using the umbrella potential every 1~MCS.
One MCS corresponds to $L^2$ attempts to flip the spin of a randomly chosen lattice site.
Our time series for $\ntil$ and $\xtil$ 
are analyzed using WHAM to evaluate $P(\ntil,\xtil)$ and $\cg(\ntil,\xtil)$.  
We estimate that the error in $\cg(\ntil,\xtil)$ is not more than $1kT$.
We exclude from the WHAM analysis any run for which the acceptance rate for the umbrella sampling is less than $0.1$, which occurs in a few cases when the local variation of 
$\cg(\ntil,\xtil)$ is very steep.

We calculate the $\cg(\ntil,\xtil)$ surface only for large values of $\ntil$ in the vicinity of $\ntil=80000$.  Also, as we will see below, the range of $\xtil$ from which we extract estimates for $S$ and $\xi$ occurs at $\xtil>0.7$.  Therefore, for the same reasons that justify Eq.~\ref{gg}, 
$\cg(\ntil,\xtil)$ as calculated here (i.e. for a system with a planar interface) estimates a section of the $G_\parallel(\ntil,\xtil)$ surface, up to an undetermined constant $C_\parallel$, where $G_\parallel$ is defined by Eq.~\ref{slab}.
We extract the cut through the $\cg(\ntil,\xtil)$ surface at fixed $\ntil_o=80000$, which is the one-dimensional function $\cg(\ntil_o,\xtil)$.  
Using the transformation given in Eqs.~\ref{ntrue} and \ref{xxx}, we then convert $\cg(\ntil_o,\xtil)$ to the 1D cut through the FES for $\cg(n,x)$ along which $n$ and $x$ vary such that $\ntil=\ntil_o$ remains constant, which we denote $G_\parallel^o$.

We fit Eq.~\ref{slab} (plus the constant $C_\parallel$) to our data for $G_\parallel^o$, 
where $S$, $\xi$ and $C_\parallel$ are the fit parameters, and where the values of $\Delta \mu_{\ca\cb}$, $\Delta \mu_{\cb\cc}$, $\sigma_{\ca\cb}$, $\sigma_{\cb\cc}$ are fixed to those reported in Ref.~\onlinecite{James:2019gi}.
We restrict the fit to data lying near $x_{\rm min}$, the value of $x$ at which the minimum of 
$G_\parallel^o$
 occurs, since this is the range of $x$ in which the stripe geometry is most stable and 
is thus where Eq.~\ref{slab} is the appropriate model of the system free energy.  Specifically, we fit using data points for which $\beta G_\parallel^o< 20$ for $x<x_{\rm min}$, and for which
$\beta G_\parallel^o< 8$ for $x>x_{\rm min}$.  For the range of $H$ studied here, we find that $x_{\rm min}$ varies from 0.88 to 0.94.

\end{document}